\documentclass[pra,twocolumn]{revtex4}
\usepackage{amsmath}
\usepackage{amsfonts}
\usepackage{amssymb}
\usepackage{graphicx}
\usepackage{sidecap}
\usepackage{caption,subcaption}
\usepackage{bm}
\DeclareMathOperator{\sech}{sech}
\begin{document}
\title{Analytic normal mode frequencies for $N$ identical particles:
The microscopic dynamics underlying the emergence and stability of excitation 
gaps from BCS to unitarity}
\author{D. K. Watson \\
University of Oklahoma \\ Homer L. Dodge Department of Physics and Astronomy \\
Norman, OK 73019}
\date{\today}

\begin{abstract}
The frequencies of the analytic normal modes for $N$ identical 
particles are studied as a function of system parameters from the weakly
interacting BCS regime to the strongly interacting unitary regime. 
The normal modes were obtained previously from a first-order $L=0$ group 
theoretic 
solution of a three-dimensional Hamiltonian with a general two-body 
interaction for confined, identical particles.
In a precursor to this study, the collective behavior of these normal modes 
was investigated
as a function of $N$ from few-body systems to many-body systems analyzing the
contribution of individual particles to the collective macroscopic
motions. A specific case, the Hamiltonian for Fermi gases in the unitary 
regime was studied in more detail. This regime is known to support
collective behavior in the form of superfluidity and has previously been 
successfully described using normal modes. Two phenomena 
that could sustain the emergence and stability of superfluid behavior were 
revealed, including the behavior of the normal mode frequencies as $N$ 
increases. In this paper, I focus on a more detailed analysis of these analytic
 frequencies,
extending my investigation to include Hamiltonians with a range
of interparticle interaction strengths from the BCS regime to the unitary 
regime  and analyzing the microscopic dynamics that lead to large gaps
at unitarity. 
The results of the current study suggest that in regimes where higher-order
effects are small, normal modes 
can be used to describe the physics of superfluidity from the
weakly interacting BCS regime with the emergence of small excitation gaps 
to unitarity with its large gaps, and can offer insight into a possible 
microscopic understanding of the behavior at unitarity.  
This approach could thus
offer an alternative to the two-body pairing models commonly used
to describe superfluidity along this transition.

\end{abstract}


\maketitle

\section{Introduction}

The evolution of collective behavior in the form of superfluidity for
systems of ultracold gaseous fermions from the BCS regime to unitarity has been
studied intensely during the last two decades since this transition was first
explored in the laboratory\cite{jin1,jin2,zwierlein1,zwierlein2,grimm1,salomon1,thomas1,hulet1,jochim1}. Typically, theoretical methods
assume that fermions are pairing into 
loosely-bound Cooper pairs in
the BCS regime to explain the emergence of superfluid 
behavior\cite{giorgini1,randeria1,leggett1,leggett2,bcs,eagles,nozieres}.  
As the interparticle interaction increases toward the
unitary regime in ultracold fermion gases, these atomic
Cooper pairs decrease in size, ultimately forming
diatomic molecules that condense in the BEC regime on the other side of
unitarity.
In materials supporting superconductivity, the emergence of Cooper 
pairs of electrons is mediated by interactions
with phonons in the underlying material creating a weak attraction that can
bind two electrons at 
long distances\cite{bcs,leggett1,leggett2,eagles,nozieres}.  
In an ultracold Fermi gas,
neutral atoms are assumed to pair at large distances due to weak
interactions when a Feshbach resonance is tuned far from resonance.
In this study, I will present an alternative possibility
to describe the transition from the weakly interacting BCS regime
to the strong interactions of the unitary regime that does not assume that
individual fermions are forming pairs. Instead, the proposed model assumes 
many-body pairing exhibited through normal modes to model the physics 
i.e. synchronized, collisionless
motion of the particles that makes it impossible to know which fermion is
paired with any other fermion.  Normal mode functions naturally
provide simple, coherent macroscopic wave functions with phase coherence 
that is maintained over the entire ensemble. The excitations between
modes define  ``quasiparticles'' of the macroscopic quantum system.
I will show how the lowest two normal
mode frequencies relevant for ultracold systems, a phonon mode and a  
particle-hole excitation mode,  
exhibit, as expected, an extremely small excitation gap in the weakly interacting
BCS regime which widens as the interaction increases reaching
 a maximum in the unitary regime.
The physics of this model from BCS to unitarity 
is a precurser to the physical pairing of 
atoms
in real space that eventually 
forms diatomic molecules in the BEC 
region.

Normal mode behavior is ubiquitous in the universe, occuring
 at all scales from the vibration of crystals\cite{NM4} to the 
oscillation of rotating stars\cite{NM8}. This universal dynamic reflects 
the widespread appearance 
of vibrational motions in nature\cite{NM1,NM2,NM3,NM4,NM5,NM6,NM7,NM8,NM9,NM10,
NM11,NM12,NM13} which can often be coupled 
into the simple collective motions of normal modes.  These collective motions
depend on the interparticle correlations of the system and thus incorporate
many-body effects into simple, dynamic motion. 
If higher order effects are small,
these collective motions are eigenfunctions of an  
approximate Hamiltonian and acquire some degree of stability 
as a function of time; thus,
a system in a single normal
mode will tend to stay in that mode unless perturbed. Normal modes 
manifest the symmetry of this underlying approximate Hamiltonian
with the possibility of offering analytic solutions to many-body problems 
and a clear physical picture of the microscopic dynamics underlying diverse
phenomena.

In an earlier paper, I studied the character of five types of normal modes
previously derived as the $L=0$ first-order analytic solutions of a
general Hamiltonian for confined, identical particles\cite{annphys}
using a perturbation formalism called symmetry-invariant perturbation
theory (SPT)\cite{FGpaper,energy,paperI,laingdensity,test,toth,JMPpaper,prl}. 
 Using the simple analytic expressions for the $N$-body 
normal mode coordinates, I 
investigated the evolution of their physical character as a function of $N$,
from few-body to many-body, and examined the motion of the individual 
particles as they 
contributed to the collective motion. Some general observations were 
made based on symmetry 
considerations and then their behavior was analyzed for a specific case,
the Hamiltonian 
of a confined system of fermions in the unitary regime which is known to 
support superfluid behavior. This study found that the 
behavior expected for few-body systems, which
 have the well known motions of molecular equivalents such as ammonia and 
methane (symmetric stretch, symmetric bend, antisymmetric stretch, 
antisymmetric bend and the opening and closing of alternative interparticle 
angles), evolves smoothly as $N$ increases to the collective motions expected 
for large $N$ ensembles (breathing, center of mass, particle-hole radial and 
angular excitations and phonon). Furthermore, the transition from few-body
behavior to large $N$ behavior was found to occur at quite low values of 
$N$ ($N \approx 10$). 
This change in character from small $N$ to large $N$
is dictated by fairly simple analytic forms that nonetheless incorporate
the intricate interplay of individual particles as they contribute to the
macroscopic motion. 
The evolution of behavior was found to be determined primarily from the 
{\it symmetry} structure 
of the Hamiltonian, and thus could be applicable to diverse
phenomena at different scales if the same symmetry is present or dominates.

The SPT formalism was developed initially for systems of  
bosons\cite{FGpaper,energy,paperI,JMPpaper,laingdensity,test,toth} and more recently 
applied to fermions\cite{prl,harmoniumpra,partition,emergence}. 
This formalism has also been tested against an exactly solvable
 model problem
of harmonically interacting particles under harmonic 
confinement\cite{test,toth,harmoniumpra,partition}. 
Exact 
agreement was found (to ten or more digits of accuracy) for the
 wave function with the exact analytic wave function obtained in an 
independent solution, confirming this general formalism 
for a fully interacting, three-dimensional 
$N$-body system\cite{test} and verifying the
analytic expressions for the normal mode frequencies and coordinates.

In the
fermion studies, the numerically demanding determination of explicitly 
antisymmetrized wave functions is avoided by using specific assignments of 
normal mode occupations to enforce the Pauli principle at
first order ``on paper'' \cite{prl,harmoniumpra,emergence,partition}. 
Ground\cite{prl}
 and excited state\cite{emergence} beyond-mean-field
energies and their degeneracies have been determined 
  allowing the construction of a partition 
function\cite{partition,emergence} and the determination of thermodynamic 
quantities\cite{partition,emergence}. Constructing the partition function 
required a large number of excited states from the 
infinite spectrum of equally spaced states,
 chosen specifically to comply 
with the enforcement of the
Pauli principle, thus connecting the Pauli principle to 
many-body interaction dynamics through the normal modes. 

The study of the thermodynamic behavior of ultracold fermions in the
unitary regime obtained quite good agreement with experimental 
data for the energy, entropy and heat capacity\cite{emergence}.
Two normal modes, selected by
the Pauli principle, were found to play a role in creating and stabilizing
the superfluid
 behavior at low temperatures, a phonon mode at ultralow temperatures
and a single particle excitation mode, i.e. a particle-hole excitation,
 as the temperature increases. 
This single-particle excitation has a much 
higher frequency  and creates a gap that stabilizes the superfluid behavior.
This normal mode description offers an interesting 
alternative to two-body pairing correlation models commonly used to describe 
superfluid regimes.

The good agreement with experiment for
thermodynamic
quantities increased the interest in investigating the
physical character of these states, originally obtained simply as a complete
basis at first order, since they
offer the possibility of acquiring physical intuition into the dynamics
of collective motion\cite{annphys} and insight into the universal behavior
of the unitary regime.

My previous study of the five types of normal mode coordinates
 looked at the evolution 
of behavior as a function 
of $N$ for only one specific case, the strongly
interacting unitary regime.  This region was chosen because of
the current experimental and theoretical interest in this regime which is
known to exhibit universal behavior and to support superfluidity with 
large excitation gaps.  My analysis of
the $N$ dependence for the unitary Hamiltonian revealed two phenomena
that have the potential to support the creation and stabilization of 
collective behavior.
First the mixing of radial and angular behavior in the normal modes is found to
limit to pure radial or pure angular behavior for very large  (or very small)
 $N$. This results
in symmetry coordinates that are eigenfunctions of an approximate Hamiltonian
governing the physics of the unitary regime, thus acquiring some amount of 
stability if the {\it symmetry} is unperturbed. Second,
for low values of $N$, the
five types of normal mode frequencies start out closer in value, 
but as $N$ increases 
these frequencies spread out creating large gaps between the values of
these five frequencies.  These gaps could provide the stability for superfluid
behavior if mechanisms to prevent the transfer of energy to
other modes exist (such as low temperatures) or could be 
engineered.

In this paper, I now extend my investigation 
to regimes other than the unitary regime, studying the evolution of the 
frequencies as a function of  the interparticle interaction strength,
$\bar{V}_0$, from the BCS regime to unitarity. I will 
focus on larger values of $N$ relevant to experimental investigations of this
transition. For this study, I have scaled the value of $\bar{V}_0$ so
$\bar{V}_0 = 1.0$ corresponds to the unitary regime which has an infinite
scattering length. The BCS regime is loosely defined as having
extremely weak interparticle interactions, e.g. $\bar{V}_0 \approx 10^{-6}$. 
This potential is briefly defined in Section~\ref{sec:SPT}
with a more detailed description in Appendix A in Ref.~\cite{annphys}.

The analytic expressions for the five types of frequencies have a complicated 
dependence on $\bar{V}_0$, both explicitly and implicitly through other
variables in the formalism that depend on $\bar{V}_0$. The goal
is to determine the interplay of various terms in the Hamiltonian
as they respond to the increase in the interparticle interaction
and affect the value of the frequencies. This understanding should offer
 insight into the microscopic dynamics that leads to large
gaps as unitarity is approached and offers
the possibility of fine tuning the
system parameters to control 
the appearance and stability of excitation gaps. 

In the remainder of this Section, I summarize the results of my investigation
and state my conclusions.
 
Section~\ref{sec:SPT} gives a brief review of the
SPT method including the derivation of the symmetry coordinates, the 
normal coordinates and the normal mode frequencies, establishing the 
necessary notation.

Section~\ref{sec:mixing} looks at the mixing coefficients (defined in 
Appendix~\ref{app:mixing}) that determine
the radial/angular mixing of the symmetry coordinates to form a normal
coordinate, extending my earlier
study in the unitary regime to regimes with weak interactions. 
Similar behavior is found for all strengths of the interparticle interaction. 
Specifically, the character of the normal modes 
 $q^{\prime [\alpha]}_{\pm}$ evolves to almost purely radial or purely
angular as $N$ increases, with very little mixing of the symmetry coordinates,
 confirming that this phenomena is 
driven by dynamics other than the universal behavior of a system at 
unitarity.  

This negligible mixing is reflected in the character of the normal mode
frequencies, which can be appropriately labelled as {\it radial} frequencies or
{\it angular} frequencies across the entire transition. 
It also has implications for the ability to tune 
these 
frequencies as well as the 
stability of collective behavior since the symmetry
coordinates are eigenfunctions of an approximate underlying Hamiltonian.
 
In Section~\ref{sec:analyticfrequencies} and 
Appendices~\ref{app:Nsectorfreq}, \ref{app:N-1sectorfreq} and 
\ref{app:N-2sectorfreq},  I analyze the analytic expressions 
for the five types of frequencies in terms of
 their dependence on $\bar{V}_0$, confirming
 that the frequencies can be characterized into two types:
{\it radial} frequencies that have a strong dependence on $\bar{V}_0$,
and {\it angular} frequencies with a weaker dependence on $\bar{V}_0$
that evolve to stable limits insensitive to changes in $\bar{V}_0$.

Section~\ref{sec:discussion} discusses the behavior of the frequencies and 
the emergence of stable gaps as a function of $\bar{V}_0$
from the BCS regime to the unitary regime.
This analysis shows the emergence of excitations gaps that increase
as $\bar{V}_0$ increases. For extremely weak interactions,
the five frequencies converge to identical values at twice the trap
frequency which results in 
infinitesimally small excitation gaps.
As $\bar{V}_0$ increases, the frequencies begin to spread out creating 
 gaps that reach a maximum at unitarity with the angular frequencies 
approaching stable limits while the radial frequencies continue to gradually
change. (These limits are derived in detail in 
Appendices~\ref{app:indeplimits} and \ref{app:limits}.)
For ultracold systems, the lowest
two frequencies are of interest, the
phonon frequency which tends to extremely small values
and the angular particle-hole frequency which limits to the trap
frequency at unitarity.  This sets up an
excitation gap that stabilizes as the unitary limit is approached. As $N$
increases, this
behavior is stabilized at smaller and smaller values of $\bar{V}_0$.
Since $\bar{V}_0$ appears as a parameter in the analytical
expressions for the frequencies,  
the evolution of these frequencies can be studied as a function of the
interparticle interaction without
intensive numerical work.

Finally in Section~\ref{sec:micro}, the microscopic dynamics underpinning
the stable limits of the angular frequencies and the emergence
of excitation gaps that could support superfluidity
 are investigated from two perspectives. First,
the relative contributions of various Hamiltonian terms to the evolving,
analytic frequencies are tracked as $\bar{V}_0$ changes. Then, 
the {\it motion} of the individual particles
in the corresponding normal mode coordinate is studied to understand
how the ensemble is rearranging on a microscopic level as interactions
turn on and collective behavior emerges. The excitation gap relevant for
ultracold Fermi gases limits to the trap frequency at unitarity, setting up
a spectrum of evenly spaced levels identical to the spectrum of the 
non-interacting regime.   This results in dynamics independent of
microscopic details of the underlying interactions consistent with the
universal behavior of the unitary regime.


\smallskip

In summary, this study of the evolution of the normal mode frequencies
from the first-order solution of the SPT equations
for confined systems of identical 
particles
as a function of $\bar{V}_0$  suggests that these normal modes are able to describe
the physics of superfluidity from the weakly-interacting BCS regime 
to the universal behavior of unitarity and to offer a view of the microscopic
dynamics without the assumption
of two-particle pairing.

\section{Symmetry-Invariant Perturbation Theory: the Derivation of the Normal Modes and their Frequencies}\label{sec:SPT}

In this Section, I summarize the development of SPT theory and
the previous derivation of the normal modes and their frequencies
that was presented in Ref.~\cite{paperI,FGpaper}, introducing the notation 
required in Sections~\ref{sec:mixing}-\ref{sec:micro} for
 the analysis of the frequencies.

The normal modes are the {\it exact} solutions at first order in inverse
dimensionality of a first principle, perturbation, many-body formalism called
symmetry-invariant perturbation theory (SPT).  This formalism  uses 
group theory to solve a fully interacting, 
 many-body, three-dimensional
Hamiltonian with a confining potential and an arbitrary interaction 
potential\cite{paperI,JMPpaper}.  Using the
symmetry of the symmetric group at large 
dimension\cite{FGpaper}, this group theoretic approach successfully rearranges
 the many-body work at each perturbation order so that an exact solution can, 
in principle, be obtained non-numerically, order-by-order, 
using group theory and graphical techniques\cite{rearrangeprl}. 
Specifically, the numerical work is rearranged
 into analytic building blocks resulting in a formulation with a complexity 
that does not scale with $N$\cite{JMPpaper, test, toth, rearrangeprl, complexity}.
Group theory is 
 used to partition the $N$ scaling problem away from the interaction 
dynamics, allowing the $N$ scaling to be treated as a  separate mathematical 
problem (cf. the Wigner-Eckart theorem). 
The exponential scaling in complexity 
is shifted from a dependence on the number of
particles, $N$, to a dependence on the order of the perturbation 
series\cite{complexity}. Exact
first-order results that contain beyond-mean-field effects for 
all values of $N$ can now be obtained from a single calculation, but 
determining higher order results becomes exponentially difficult. 
To minimize the work needed
for new calculations, the analytic building blocks
have been calculated and stored\cite{epaps}. 
Strongly interacting systems can be studied since the perturbation does not 
involve the strength of the interaction.

The Schr\"odinger equation in $D$ dimensions
 is defined in Cartesian coordinates
for $N$ interacting particles by:
\begin{equation} \label{eq:generalH} 
H \Psi  =  \left[ \sum\limits_{i=1}^{N} h_{i} +
\sum_{i=1}^{N-1}\sum\limits_{j=i+1}^{N} g_{ij} \right] \Psi = E
\Psi \,,  
\end{equation} 
\begin{equation} \label{eq:generalH1} 
\begin{array}{rcl}
h_{i} & = & -\frac{\hbar^2}{2
m_{i}}\sum\limits_{\nu=1}^{D}\frac{\partial^2}{\partial
x_{i\nu}^2} +
V_{\mathtt{conf}}\left(\sqrt{\sum\nolimits_{\nu=1}^{D}x_{i\nu}^2}\right)
\,,  \\
g_{ij} & = & V_{\mathtt{int}}\left(\sqrt{\sum\nolimits_{\nu=1}^{D}\left(x_{i\nu}-x_{j\nu}
\right)^2}\right),  
\end{array}
\end{equation}
\noindent where $h_{i}$ is the single-particle Hamiltonian,
$g_{ij}$ is a two-body interaction potential,
 $x_{i\nu}$ is the $\nu^{th}$ Cartesian component
of the $i^{th}$ particle,  and 
$V_{\mathtt{conf}}$ is a spherically-symmetric confining 
potential\cite{paperI,FGpaper,JMPpaper}. 
The  Schr\"odinger equation is transformed from Cartesian coordinates
to internal 
coordinates using: 
\begin{equation}\label{eq:int_coords}
\renewcommand{\arraystretch}{1.5}
\begin{array}{rcl} r_i & = &\sqrt{\sum_{\nu=1}^{D} x_{i\nu}^2}\,, \;\;\; (1 \le i \le
N)\,,
\;\;\; 
\\ \gamma_{ij} & = & cos(\theta_{ij})=\left(\sum_{\nu=1}^{D}
x_{i\nu}x_{j\nu}\right) / r_i r_j\,,
\end{array}
\renewcommand{\arraystretch}{1}
\end{equation}
\noindent $(1 \le i < j \le N)$\,, which are the $D$-dimensional scalar radii $r_i$ of the $N$
particles from the center of the confining potential and the
cosines $\gamma_{ij}$ of the $N(N-1)/2$ angles between the radial
vectors.

The first-order derivatives are removed using a similarity 
transformation\cite{avery}, and dimensionally-scaled oscillator units 
are defined  with a scale factor, $\kappa(D)$, that regularizes
the large-dimension limit of the Schr\"odinger equation.
Substituting the scaled variables, $\bar{r}_i = r_i/\kappa(D), \,\,
 \bar{E} = \kappa(D) E$ and $\bar{H} = \kappa(D)H$, 
into the similarity-transformed Schr\"odinger equation gives:
\begin{equation} \label{eq:scaleH_BEC}
\bar{H} \Phi =
\left(\delta^2\bar{T}+\bar{U}+\bar{V}_{\mathtt{conf}}+\bar{V}_{\mathtt{int}}\right)\Phi = \bar{E}\, \Phi\,.
\end{equation}
where
\begin{equation}\label{eq:Tbar_BEC}
\begin{split}
\bar{T} &  = \sum\limits_{i=1}^{N} \Bigl( -\frac{1}{2}\frac{\partial^2}
{{\partial \bar{r}_i}^2} \\
& - \frac{1}{2 \bar{r}_i^2}\sum\limits_{j\not=i}\sum\limits_{k\not=i}
\frac{\partial}{\partial\gamma_{ij}}(\gamma_{jk}-\gamma_{ij}\gamma_{ik})
\frac{\partial}{\partial\gamma_{ik}} \Bigr) \,,
\end{split}
\end{equation}
\begin{eqnarray}
\label{eq:Ubar_BEC}
\bar{U}&=&\sum\limits_{i=1}^{N}\left(\frac{\delta^2N(N-2)+(1-\delta(N+1))^2 \left(\frac{\Gamma^{(i)}}{\Gamma}\right)}{8 \bar{r}_i^2}\right) \,,
\\
\bar{V}_{\mathtt{conf}}&=&\sum\limits_{i=1}^{N}\frac{1}{2}\bar{r}_i^2
\\
\bar{V}_{\mathtt{int}}&=&  \frac{\bar{V}_{0}}{1-3b\delta}
\sum\limits_{i=1}^{N-1}\sum\limits_{j=i+1}^{N}
\left(1-\tanh \Theta_{i,j} \right) \,,
\end{eqnarray}
\noindent  $\hbar=m=1$, $\delta=1/D$ is the perturbation 
parameter, $\Gamma$ is the Gramian determinant with elements $\gamma_{ij}$ (see
Appendix D in Ref~\cite{FGpaper}), and
$\Gamma^{(i)}$ is the determinant with the row
and column of the $i^{th}$ particle deleted.  
The barred quantities have been scaled by $\kappa(D)$. In the
expression for $\bar{V}_{\mathtt{int}}$, the constant $b$ in the denominator is 
chosen so the potential yields an infinite scattering length for the
unitary regime with $\bar{V}_0 = 1.0$.  To evolve toward the weaker 
interactions of the BCS regime, $\bar{V}_0$ is scaled to smaller
values. The argument $\Theta_{ij}$ is defined as
\begin{equation}\label{eq:thetaij_scaled}
\Theta_{ij}=\frac{\bar{c}_0}{1-3\delta}
\left(\frac{\bar{r}_{ij}}{\sqrt{2}}-\bar{\alpha}-3\delta\left(\bar{R}-\bar{\alpha} \right) \right)
\,,
\end{equation}
where $\bar{r}_{ij}$ is the interatomic separation,
\begin{equation}\label{eq:rbarij}
\bar{r}_{ij}={\sqrt{{\bar{r}_i}^2+
{\bar{r}_j}^2-2\bar{r}_i\bar{r}_j\gamma_{ij}}}\,,
\end{equation} 
$\bar{R}$ is the range of the square-well potential in dimensionally-scaled oscillator units, and $\bar{\alpha}$ is a constant which softens the
potential as $D \rightarrow \infty$.

Taking the $D\to\infty$ limit, the second derivative terms drop out
 yielding a static zeroth-order
problem with an effective potential, $\bar{V}_{\mathtt{eff}}$:
\begin{eqnarray}\label{eq:veff_BEC}
\bar{V}_{\mathtt{eff}}(\bar{r},\gamma;\delta)&=&
\sum\limits_{i=1}^{N}\left(\bar{U}(\bar{r}_i;\delta)
+\bar{V}_{\mathtt{conf}}(\bar{r}_i;\delta)\right) \nonumber \\
&& +\sum\limits_{i=1}^{N-1}\sum\limits_{j=i+1}^{N}
\bar{V}_{\mathtt{int}}(\bar{r}_i,\gamma_{ij};\delta)\,.
\end{eqnarray}
The minimum
of this effective potential yields an infinite-dimensional maximally-symmetric
structure 
with all radii, $\bar{r}_i$, and angle
 cosines, $\gamma_{ij}$, of the particles equal, i.e. when $D\to\infty$, 
$\bar{r}_{i}=\bar{r}_{\infty} \;\; (1 \le i \le N)$ and
$\gamma_{ij}={\gamma}_\infty \;\; (1 \le i < j \le N)$.
The values of these parameters are determined by 
two  minimum conditions:
\begin{equation}\label{eq:minimum2}
\left(\frac{\partial \bar{V}_\text{eff}}
{\partial\bar{r}_{i}}\right)\Biggr|_{\infty}=0, \,\,\,\,\,\,\,\,\,\,\,\,
\left(\frac{\partial \bar{V}_\text{eff}}
{\partial\gamma_{ij}}\right)\Biggr|_{\infty}= 0\,.
\end{equation}
\noindent Substituting the above definition of $\bar{V}_\textrm{eff}$, 
two equations in $\bar{r}_\infty$ and $\gamma_\infty$ are obtained 
which yield:
\begin{equation}\label{eq:rinfeq}
 \bar{r}_\infty=\frac{1}{\sqrt{2} \sqrt{1+(N-1) \gamma _{\infty }}}\,,
\end{equation}
\noindent while $\gamma_{\infty}$ can be solved from 
the transcendental equation:
\begin{equation}\label{eq:gammaeq0bec}
   \frac{\gamma _{\infty } \left(2+(N-2) \gamma _{\infty
   }\right)}{\left(1-\gamma _{\infty }\right){}^{3/2}
   \sqrt{1+(N-1) \gamma _{\infty }}}
   +
   \bar{V}_0\,\text{sech}^2\left(\Theta _{\infty}\right)\Theta _{\infty }'
   =
   0\,.
\end{equation}
\noindent In the large-$D$ limit ($\delta \rightarrow 0$), 
the argument $\Theta_{ij}$ becomes
\begin{equation}
\label{eq:thetainf2}
\Theta_\infty
=\Theta_{ij}\Biggr|_{\infty}
=
\bar{c}_0 \left(\sqrt{1-\gamma _{\infty }}
   \,\bar{r}_{\infty }-\bar{\alpha }\right) \,,
\end{equation}
\noindent The zeroth-order energy at this minimum, 
$\bar{E}_\infty =\bar{V}_{\mathtt{eff}}(\bar{r}_\infty)$ provides the starting point for the $1/D$
expansion.
A position
vector of the $N(N+1)/2$ internal coordinates is defined as:
\begin{equation}\label{eq:ytranspose}
\begin{array}[t]{l} {\bar{\bm{y}}} = \left( \begin{array}{c} \bar{\bm{r}} \\
\bm{\gamma} \end{array} \right) \,, \;\;\; \mbox{where} \;\;\; \\
\mbox{and} \;\;\; \bar{\bm{r}} = \left(
\begin{array}{c}
\bar{r}_1 \\
\bar{r}_2 \\
\vdots \\
\bar{r}_N
\end{array}
\right) \,. \end{array} 
\bm{\gamma} = \left(
\begin{array}{c}
\gamma_{12} \\ \cline{1-1}
\gamma_{13} \\
\gamma_{23} \\ \cline{1-1}
\gamma_{14} \\
\gamma_{24} \\
\gamma_{34} \\ \cline{1-1}
\gamma_{15} \\
\gamma_{25} \\
\vdots \\
\gamma_{N-2,N} \\
\gamma_{N-1,N} \end{array} \right) \,,
\end{equation}

\noindent The substitutions: 
$\bar{r}_{i} = \bar{r}_{\infty}+\delta^{1/2}\bar{r}'_{i}$
and $\gamma_{ij} =
{\gamma}_{\infty}+\delta^{1/2}{\gamma}'_{ij}$ set up a
power series in $\delta^{1/2}$ about
the $D\to\infty$ symmetric minimum.
%


The first-order, $\delta=1/D$, equation is a harmonic problem,
which is solved exactly and analytically 
by obtaining the $N$-body normal modes of the system.
 The first-order  Hamiltonian, $\bar{H}_1$,
 is defined in terms of constant matrices 
$\bm{G}$ and ${\bf F}$ that are evaluated at the large
dimension limit:

\begin{equation} \label{eq:Gham}
\bar{H}_1=-\frac{1}{2} {\partial_{\bar{y}'}}^{T} {\bm G}
{\partial_{\bar{y}'}} + \frac{1}{2} \bar{\bm{y}}^{\prime T} {\bm
F} {{\bar{\bm{y}}'}} + v_o \,.
\end{equation}

\noindent where $G$ involves kinetic energy terms, $F$ involves derivatives
 of the effective potential, and $v_o$ is a constant\cite{FGpaper}.
The FG matrix method\cite{dcw} is used to obtain the normal-mode
frequencies and the harmonic-order energy correction\cite{FGpaper}. A
review of the FG matrix method is presented in Appendix A of 
Ref.~\cite{FGpaper}.
$N(N+1)/2$ frequencies, $\bar{\omega}$, are obtained from the roots of
the FG equation, 
however only five roots are distinct due to the
large degeneracy of the
frequencies reflecting the very high degree of symmetry manifested in
the $\bm{F}$\,, $\bm{G}$\,,
and $\bm{FG}$ matrices.
The elements of these matrices are evaluated for the large dimension,
maximally-symmetric structure with a single value for all radii 
$\bar{r}_\infty$ and 
angle cosines, ${\gamma}_{\infty}$. Thus these matrices are
invariant under the $N!$ operations of particle interchanges 
effected by the symmetric group, $S_N$, which results in the highly 
degenerate eigenvalues.
These matrices do not connect
subspaces belonging to different irreducible representations of 
$S_N$\cite{WDC,hamermesh},
thus 
the normal coordinates must transform under irreducible
representations of $S_N$\,.

There are a total of five irreducible
representations: two $1$-dimensional
irreducible representations, one radial and one angular,
 labelled by the partition $[N]$,
 two  $(N-1)$-dimensional irreducible
representations, one radial and one angular, labelled by the 
partition $[N-1, \hspace{1ex} 1]$,
and one angular $N(N-3)/2$-dimensional irreducible representation labelled
by the partition $[N-2, \hspace{1ex} 2]$.
These representations are given shorthand labels:
 ${\bf 0}^-$, ${\bf 0}^+$, ${\bf 1}^-$, 
${\bf 1}^+$, and ${\bf 2}$ respectively,
(see Refs.~\cite{paperI,JMPpaper}).
Thus the energy through first-order in $\delta=1/D$
can be
written in terms of the five distinct normal mode 
frequencies\cite{FGpaper,loeser} as:

\begin{equation}
\overline{E} = \overline{E}_{\infty} + \delta \Biggl[
\sum_{\renewcommand{\arraystretch}{0}
\begin{array}[t]{r@{}l@{}c@{}l@{}l} \scriptstyle \mu = \{
  & \scriptstyle \bm{0}^\pm,\hspace{0.5ex}
  & \scriptstyle \bm{1}^\pm & , 
  &  \,\scriptstyle \bm{2}   \scriptstyle  \}
            \end{array}
            \renewcommand{\arraystretch}{1} }
(n_{\mu}+\frac{1}{2} d_{\mu})
\bar{\omega}_{\mu} \, + \, v_o \Biggr] \,. \label{eq:E1}
\end{equation}

\noindent where  
 $\mu$ is a label which
runs over the five types of normal modes, ${\bf 0}^-$\,, ${\bf
0}^+$\,, ${\bf 1}^-$\,, ${\bf 1}^+$\,, and ${\bf 2}$\,, (irrespective of
 the particle number, see Ref.~\cite{FGpaper} and Ref.[15]
in \cite{paperI}),
$n_{\mu}$ is the total quanta in the normal mode
with frequency $\bar{\omega}_{\mu}$;
 and $v_o$ is a constant (defined in 
Ref.~\cite{FGpaper}, Eq.(125)).
The
multiplicities of the five roots are:
$d_{{\bf 0}^+} = 1, \hspace{1ex} d_{{\bf 0}^-} = 1,\;
d_{{\bf 1}^+} = N-1,\;  d_{{\bf 1}^-} = N-1,\;
d_{{\bf 2}} = N(N-3)/2$.

We define the symmetry coordinate vector, $S$, as:

\begin{equation}\label{eq:trial}
\bm{S} = \left( \begin{array}{l} {\bm{S}}_{\bar{\bm{r}}'}^{[N]} \\
{\bm{S}}_{\overline{\bm{\gamma}}'}^{[N]} \\
{\bm{S}}_{\bar{\bm{r}}'}^{[N-1, \hspace{1ex} 1]} \\
{\bm{S}}_{\overline{\bm{\gamma}}'}^{[N-1, \hspace{1ex} 1]} \\
{\bm{S}}_{\overline{\bm{\gamma}}'}^{[N-2, \hspace{1ex} 2]}
\end{array} \right) =
\left( \begin{array}{l}W_{\bar{\bm{r}}'}^{[N]} \, \bar{\bm{r}}' \\ 
W_{\overline{\bm{\gamma}}'}^{[N]} \, \overline{\bm{\gamma}}' \\
W_{\bar{\bm{r}}'}^{[N-1, \hspace{1ex} 1]} \bar{\bm{r}}'\\
W_{\overline{\bm{\gamma}}'}^{[N-1, \hspace{1ex} 1]} \,
\overline{\bm{\gamma}}' \\
W_{\overline{\bm{\gamma}}'}^{[N-2, \hspace{1ex} 2]} \,
\overline{\bm{\gamma}}' \end{array} \right) \,,
\end{equation}
where the $W_{\bar{\bm{r}}'}^{[\alpha]}$ and the $ W_{\bar{\bm{\gamma}}'}^{[\alpha]}$
are transformation matrices. 
This is shown in Ref.~\cite{paperI} using the theory of group characters 
to decompose
$\bar{\bm{r}}'$ and $\overline{\bm{\gamma}}'$ into basis functions that 
transform under these five irreducible representations of $S_N$.
%
%

The $\bm{FG}$ method is applied using these symmetry
coordinates to determine the eigenvalues, 
$\lambda_\alpha = {\bar{\omega}_\alpha}^2$, 
frequencies, $\bar{\omega}_\alpha$, and normal modes, ${\bm{q}'}^{[\alpha]}$,
 of the system:

\begin{equation} \label{eq:qnpfullexp}
{\bm{q}'}_{\pm}^{[N]}  = c_{\pm}^{[N]} \left(
\cos{\theta^{[N]}_{\pm}} \, [{\bm{S}}_{\bar{\bm{r}}'}^{[N]}]
\, + \, \sin{\theta^{[N]}_{\pm}} \,
[{\bm{S}}_{\overline{\bm{\gamma}}'}^{[N]}] \right)  
\end{equation}

\begin{equation} \label{eq:qnpfullexp1}
\begin{split}
{\bm{q}'}_{\xi\pm}^{[N-1,1]} & = c_{\pm}^{[N-1,1]} \Bigl(
\cos{\theta^{[N-1,1]}_{\pm}} [{\bm{S}}_{\bar{\bm{r}}'}^{[N-1,1]}]_{\xi}  \\
& \,\,\,\,\,\,\,\,\,\,\,\,\,\,\,\,\,\,\,\,\,
+ \sin{\theta^{[N-1,1]}_{\pm}}[{\bm{S}}_{\overline{\bm{\gamma}}'}^{[N-1,1]}]_{\xi}
 \Bigr) \, 
\end{split}
\end{equation}
\noindent for the $\alpha=[N]$ and $[N-1, \hspace{1ex} 1]$ sectors,
 $1 \leq \xi \leq N-1$ and
\begin{equation} \label{eq:qnm2fullexp}
{\bm{q}'}^{[N-2, \hspace{1ex} 2]} = c^{[N-2, \hspace{1ex} 2]}
{\bm{S}}_{\overline{\bm{\gamma}}'}^{[N-2, \hspace{1ex} 2]} \,
\end{equation}
for the $[N-2, \hspace{1ex} 2]$ sector.

From Eqs.~(\ref{eq:qnpfullexp}) and (\ref{eq:qnpfullexp1}) above, 
the symmetry coordinates 
in the 
$[N]$ and $[N-1,1]$ sectors are mixed to form a normal coordinate.
Thus, depending on the value of the mixing angles, the normal modes,
which are the  eigenfunctions at first order of the
Schr\"odinger equation will have mixed radial and angular behavior 
in the $[N]$ and 
$[N-1,1]$ sectors.
The $[N-2,2]$ normal modes have entirely angular behavior since there are
no  $\bar{\bm{r}}'$ symmetry coordinates in this sector and so no mixing. 
The value of the mixing angles and
thus the extent of radial/angular mixing in a normal coordinate
 depends, of course, on the Hamiltonian terms at this first perturbation
order.

\section{Mixing coefficients as a function of $N$ and $\bar{V}_0$.}\label{sec:mixing}

The mixing coefficients that determine the radial/angular mixing in the
normal modes for the $[N]$ and $[N-1,1]$ sectors are              
defined in Appendix~\ref{app:mixing} and
have a complicated dependence on $N$ and $\bar{V}_0$ that originates in the 
Hamiltonian terms at first order. 
In particular, these coefficients have some explicit dependence from
 the {\it symmetry} present in the first-order Hamiltonian as well as
dependence from the $F$ and $G$ elements for a particular Hamiltonian.

As shown in Appendix~\ref{app:mixing}, there are three layers of analytic 
expressions
that can bring in $N$ and/or $\bar{V}_0$ dependence.
When these mixing coefficients were plotted for the {\it unitary} (large
$\bar{V}_0$) Hamiltonian
as a function of $N$ in a recent study, 
the character of the normal modes was found
to evolve to a pure radial or pure angular symmetry coordinate for $N \ge 200$
i.e. no mixing for $N>>1$. (This was also true for very small $N$ 
e.g. $N \le 10$ which are not being studied in this work.)

A bit of inspection revealed that this behavior was being dictated 
to a large extent by the explicit $N$
dependence in the expressions
for $\cos\theta^{[\alpha]}_{\pm}$ and $\sin\theta^{[\alpha]}_{\pm}$ 
(Eqs.~(\ref{eq:cos0p})-(\ref{eq:sin1m})).
These
expressions depend on the symmetry of the first-order Hamiltonian, not
the specific details of the potential.
The position and shape of the crossover is influenced by the other 
sources of $N$ and $\bar{V}_0$ dependence that originate in the specific 
Hamiltonian.

The emergence of pure symmetry coordinates for 
large $N$ has implications for the stability of collective behavior 
since the symmetry
coordinates are eigenfunctions of an approximate underlying Hamiltonian
and thus have some degree of stability unless the system is perturbed. 
In addition, this means that the {\it frequencies},
 $\bar{\omega}_{{[N]}^{\pm}}$ and $\bar{\omega}_{{[N-1,1]}^{\pm}}$,
 associated with these normal modes
should reflect pure radial or pure angular character for large $N$.

In the current study, I now extend this earlier study to regimes other than the 
unitary regime, investigating whether this emergence of pure symmetry
character in the normal modes and their frequencies is unique to the
unitary regime or is driven by dynamics common to regimes throughout
the transition from BCS to unitarity.

%
In Figs.~(\ref{fig:trialone})-(\ref{fig:trialfour}),
I show the behavior of the mixing coefficients as a function of
$N$ for a system of identical fermions in the weakly interacting
BCS regime, 
plotting the square of the mixing coefficients, 
${|\cos\theta^{[\alpha]}_{\pm}|^2}$ and ${|\sin\theta^{[\alpha]}_{\pm}|^2}$ 
which gives the probability associated with each symmetry
coordinate, $[{\bm{S}}_{\bar{\bm{r}}'}^{[\alpha]}]$ or
$[{\bm{S}}_{\overline{\bm{\gamma}}'}^{[\alpha]}] $,
 in the expression for the 
normal modes, $q^{\prime [\alpha]}_{\pm}$.

\twocolumngrid

\begin{figure}
\centering
\begin{subfigure}[h]{0.51\textwidth}
\includegraphics[scale=0.65]{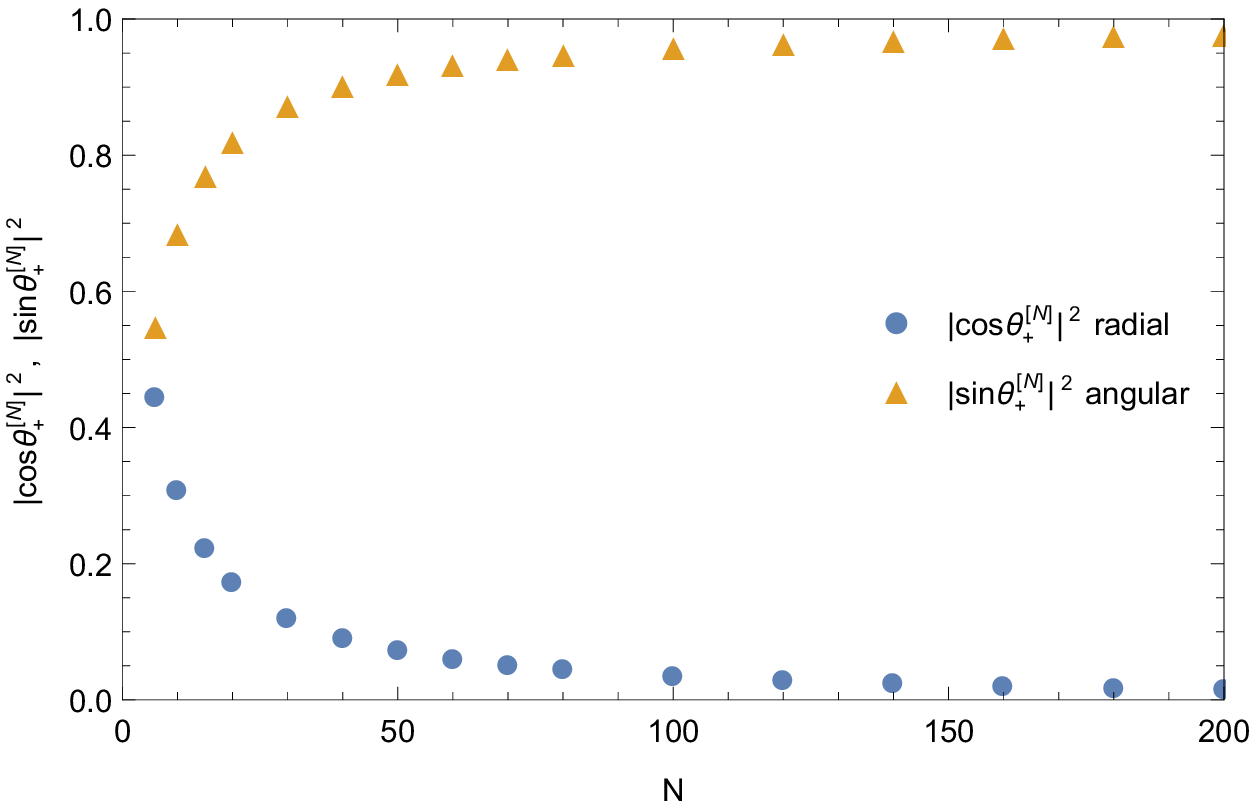}
\caption{a. $|\cos\theta^{[N]}_+|^2$ 
and $|\sin\theta^{[N]}_+|^2$ for the normal mode $q^{\prime [N]}_+$.}
\label{fig:trialone}
\end{subfigure}
\begin{subfigure}[h]{0.51\textwidth}
\includegraphics[scale=0.65]{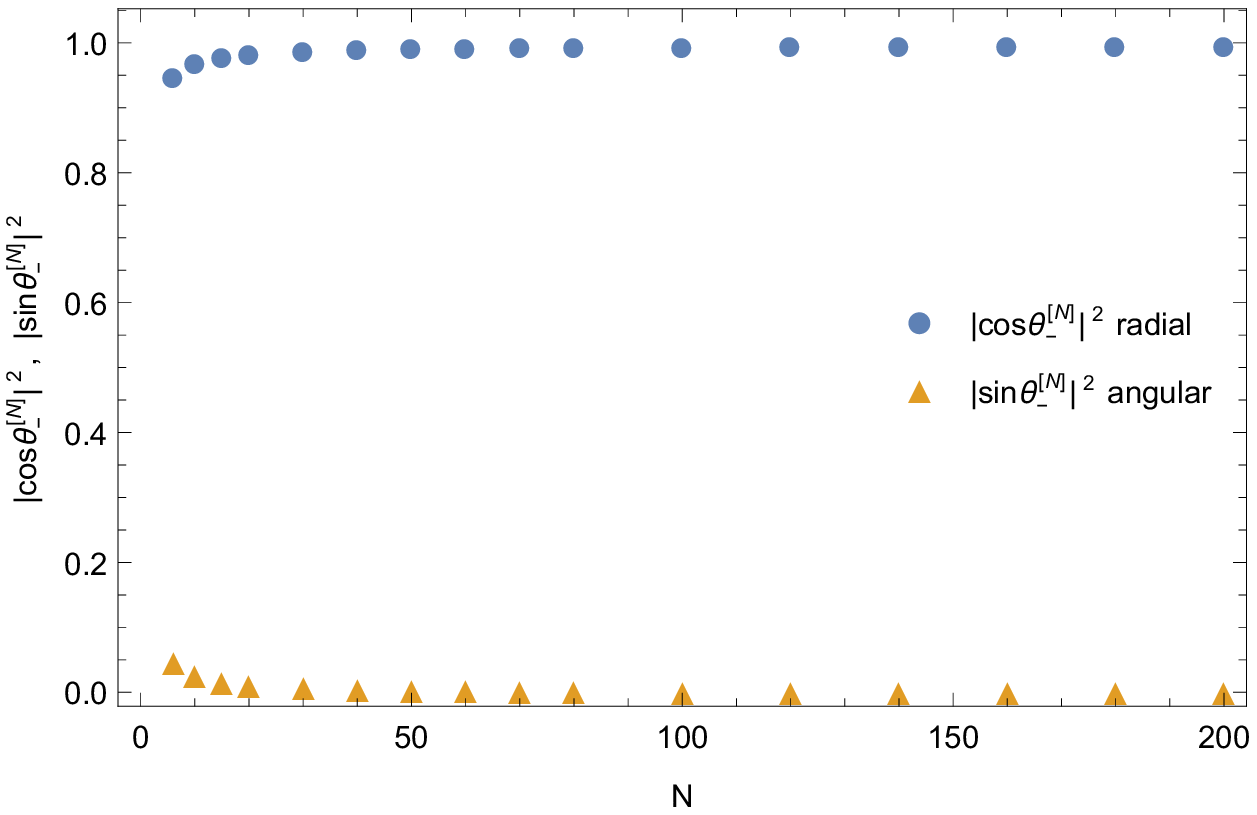}
\caption{b. $|\cos\theta^{[N]}_-|^2$ 
and $|\sin\theta^{[N]}_-|^2$ for the normal mode $q^{\prime [N]}_-$.}
\label{fig:trialtwo}
\end{subfigure}
\begin{subfigure}[h]{0.51\textwidth}
\includegraphics[scale=0.65]{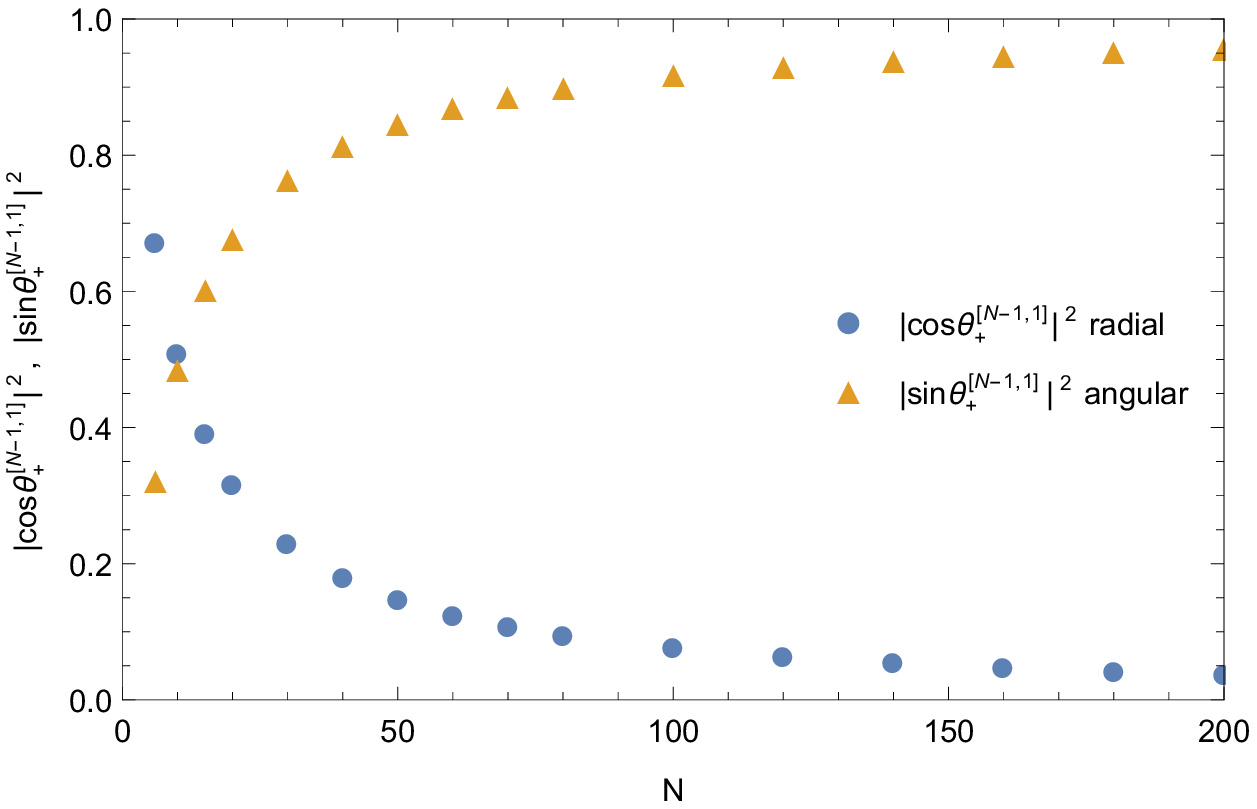}
\caption{c. $|\cos\theta^{[N-1,1]}_+|^2$ 
and $|\sin\theta^{[N-1,1]}_+|^2$ for $q^{\prime [N-1,1]}_+$.}
\label{fig:trialthree}
\end{subfigure}
\begin{subfigure}[h]{0.51\textwidth}
\includegraphics[scale=0.65]{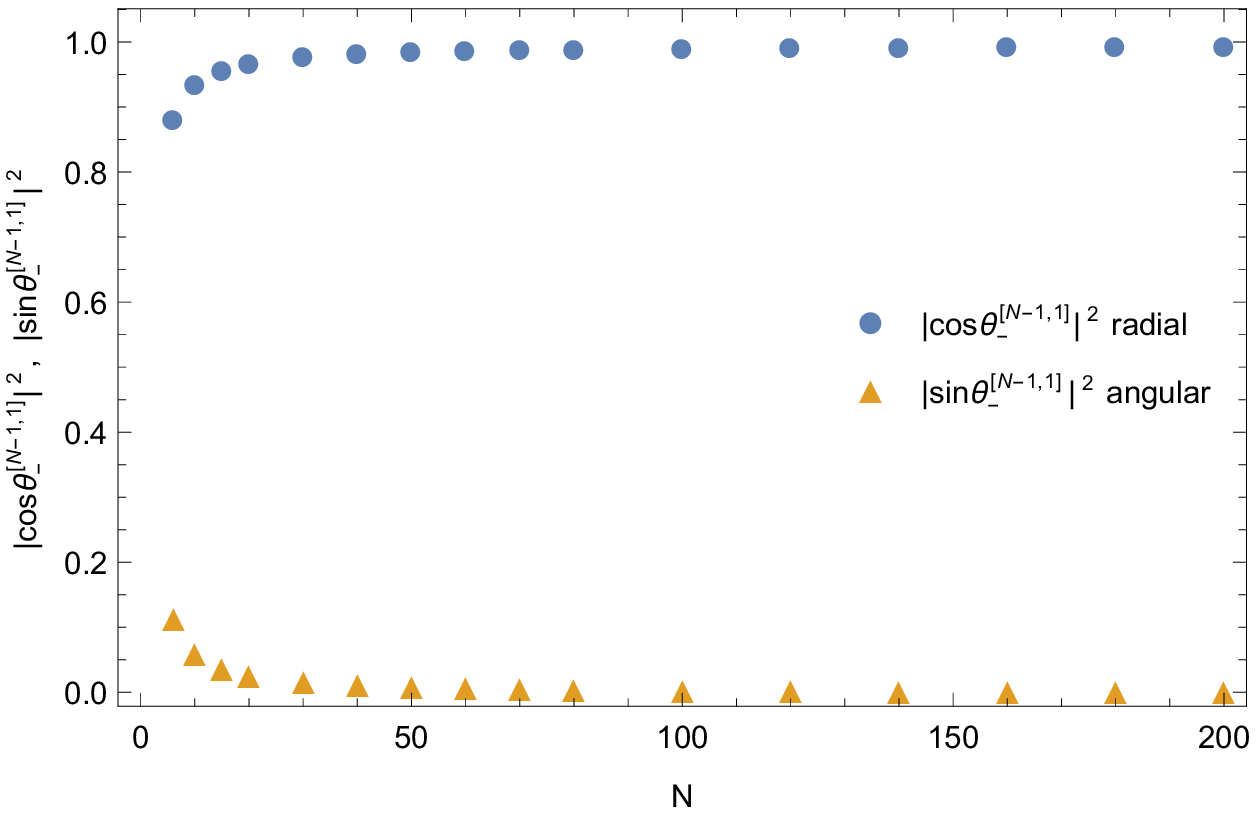}
\caption{d. $|\cos\theta^{[N-1,1]}_-|^2$ 
and $|\sin\theta^{[N-1,1]}_-|^2$ for $q^{\prime [N-1,1]}_-$}
\label{fig:trialfour}
\end{subfigure}
\setcounter{figure}{0}
\caption{The square of the mixing coefficients as a function of N for the 
weakly interacting BCS regime.}
\label{fig:one2}

\end{figure}

The plots show that the character of the normal modes 
 $q^{\prime [\alpha]}_{\pm}$ evolves to almost purely radial or purely
angular, as $N$ increases with very little mixing of the symmetry coordinates.
This happens in this weakly interacting regime at even lower values of $N$
than in the unitary regime, thus confirming that this phenomena is 
driven by dynamics other than the universal behavior of a system at 
unitarity.  This also validates the decision to designate
the  $[N]$ and $[N-1,1]$ sector normal mode frequencies 
for the typical many-body ensemble sizes studied in the
laboratory, as either a {\it radial} frequency or an
{\it angular} frequency instead of having mixed radial/angular character.
Inspecting the plots reveals that
$\bar{\omega}_{{0}^+}$ and 
$\bar{\omega}_{{1}^+}$ are {\it angular} frequencies and $\bar{\omega}_{{0}^-}$
and $\bar{\omega}_{{1}^-}$ are {\it radial} frequencies.  (This designation 
holds 
over the entire range of interparticle interaction strengths until the
systems are approaching the unitary regime where the large value of $\bar{V}_0$
results in a crossing of the character at which point the labels are switched.)

\section{The Analytic Expressions for the Five Normal Mode
Frequencies}
\label{sec:analyticfrequencies}

In this section, I analyze the analytic expressions for the frequencies,
 investigating
the differences between the radial, $\bar{\omega}_{0^{-}}$, 
$\bar{\omega}_{1^{-}}$,
and angular, $\bar{\omega}_{0^{+}}$, $\bar{\omega}_{1^{+}}$, frequencies 
 in the $[N]$ and $[N-1,1]$ 
sectors, as well as studying the angular 
frequency $\bar{\omega}_2$ in the
$[N-2,2]$ sector. The radial 
frequencies depend strongly on the  interparticle interaction potential,
$\bar{V}_0$,
 while the angular frequencies which are comprised primarily from centrifugal 
potential terms have a weaker dependence on $\bar{V}_0$.
Thus all the frequencies will respond to tuning the interaction strength
in the laboratory using  a Feshbach resonance.
 
Analytic expressions for the $N$-body normal mode frequencies were derived in 
Ref.~\cite{FGpaper} using a
method  outlined
in Appendices B and C of that paper which derives analytic formulas for the 
roots, $\lambda_{\mu}$, of 
the $FG$ secular equation.
The  normal-mode vibrational
frequencies, $\bar{\omega}_{\mu}^2$, are related to the roots $\lambda_{\mu}$ of
${\bf FG}$ by:
\begin{equation}\label{eq:omega_p}
\lambda_{\mu}=\bar{\omega}_{\mu}^2,
\end{equation}
The two  frequencies associated with the $\lambda_{0}$ roots of 
multiplicity one are of the form:
\begin{equation}\label{eq:omega0}
\bar{\omega}_{{0}^{\pm}}=\sqrt{\eta_0 \mp
\sqrt{{\eta_0}^2-\Delta_0}},
\end{equation}
\noindent where:
\begin{equation}\label{eq:lam0defs}
\begin{split}
\eta_0 &= \frac{1}{2} \Biggl[a+(N-1)b+g+2(N-2)h \\
& \,\,\,\,\,\,\,\,\,\,\,\,\,\,\,\,\,\,\,\,\,\,\,\,\,\,\,\,\,\,\,\,\, +\frac{(N-2)(N-3)}{2}\iota \Biggr] \\ 
\Delta_0 &= (a+(N-1)b)\left[g+2(N-2)h+\frac{(N-2)(N-3)}{2}\iota\right]  \\
&\,\,\,\,\,\,\,\,\,\,\,\,\,\,\,\,\,-\frac{N-1}{2}(2c+(N-2)d)(2e+(N-2)f).
\end{split}
\end{equation}
For the two $N-1$ multiplicity roots, the frequencies are:
\begin{equation}\label{eq:omega1}
\bar{\omega}_{{1}^{\pm}}=\sqrt{\eta_1 \mp
\sqrt{{\eta_1}^2-\Delta_1}},
\end{equation}
\noindent where $\eta_1$ and $\Delta_1$ are given by:
\begin{eqnarray}\label{eq:lam1defs}
\eta_1 &=& \frac{1}{2}\left[a-b+g+(N-4)h-(N-3)\iota\right] \nonumber \\
\Delta_1 &=&-(N-2)(c-d)(e-f)+(a-b) \nonumber \\
 &&\times\left[g+(N-4)h-(N-3)\iota\right].
\end{eqnarray}
The frequency ${\bar{\omega}}_2$, associated with the root $\lambda_2$ of
multiplicity $N(N-3)/2$ is given by:
\begin{equation}\label{eq:omega2}
\bar{\omega}_2=\sqrt{g-2h+\iota}.
\end{equation}
The quantities $a,b,c,d,e,f,g,h, \mbox{ and } \iota$, are defined
in Appendix~\ref{app:FG}  in terms of the $F$ and $G$ elements of the
first-order Hamiltonian.

\smallskip

\subparagraph{Explicit $\bar{V}_0$ dependence in the analytic expressions for 
the frequencies.} Analogous to the mixing coefficients, there are
three layers of analytic expressions that define the frequencies:
the expressions for 
$ \bar{\omega}_{{0}^{\pm}}$, $\bar{\omega}_{{1}^{\pm}}$ and $\bar{\omega}_2 $
 in Eqs.~(\ref{eq:omega0})-(\ref{eq:omega2}) above, 
the expressions for $a,b,c,d,e,f,g,h$ and $\iota$ given 
in Appendix~\ref{app:FG},
and the expressions for the $F$ and $G$ elements of  Eq.~(\ref{eq:Gham})
 which are also given in  Appendix~\ref{app:FG}. 
The expressions for
$ \bar{\omega}_{{0}^{\pm}}$, $\bar{\omega}_{{1}^{\pm}}$ and $\bar{\omega}_2 $
 do not contain $\bar{V}_0$ explicitly, however
the next layer, which involves the quantities 
$a,b,c,d,e,f,g,h, \mbox{ and } \iota$, does have explicit dependence 
on $\bar{V}_0$, as well as the third layer involving the $F$ and $G$ elements   
of the first-order Hamiltonian.

\smallskip

\subparagraph{Implicit dependence of the frequencies on 
$\bar{V}_0$ through the variables $\bar{r}_{\infty}$ , $\gamma_{\infty}$,
$\tanh\Theta_{\infty}$ and $\sech^2\Theta_{\infty}$.}
The frequencies have some implicit dependence
on $\bar{V}_0$ from the variables $\bar{r}_{\infty}$ and 
$\gamma_{\infty}$ whose values
 are obtained as roots of transcendental equations 
that involve  
$\bar{V}_0$ (see Eqs.~(\ref{eq:minimum2})-(\ref{eq:gammaeq0bec})). 
The values of
$\bar{r}_{\infty}$ and $\gamma_{\infty}$ are also used to determine 
$\tanh\Theta_{\infty}$ and $\sech^2\Theta_{\infty}$.
This implicit dependence
on the interparticle interaction strength through
the solution of  transcendental equations complicates understanding 
the dependence of the frequencies on $\bar{V}_0$
solely by analytic means, however it is possible with a little
numerical work to understand how  $\bar{V}_0$ is implicitly 
affecting the frequencies through these variables.

\smallskip

\subsection{The $[N]$ sector frequencies, $\bar{\omega}_{0^{\pm}}$.}

The normal mode frequencies in the $[N]$ sector, $\bar{\omega}_{0^{+}}$ and
$\bar{\omega}_{0^{-}}$, are associated with the angular center of mass mode
and the radial breathing mode respectively. These two frequencies are the 
largest frequencies 
of the five normal modes and so do not come into play in providing an
excitation gap in ultracold regimes.  It is interesting to analyze the
relative contributions of the interaction potential, $\bar{V}_0$, versus 
terms originating in the
kinetic energy for these $[N]$ sector frequencies.
 The breathing frequency is expected to depend strongly
on the strength of $\bar{V}_0$ as the particles
spread out and then move back in toward the minimum of the effective
potential. For the center of mass
frequency, the dependence on $\bar{V}_0$ should
drop out of the final simplified analytic expression since the center of
mass mode is independent of interparticle interactions.
In Appendix~\ref{app:Nsectorfreq}, I demonstrate how the quadratic formula for
$\bar{\omega}_{0^{+}}$ and $\bar{\omega}_{0^{-}}$ results in one frequency,
$\bar{\omega}_{0^{-}}$, the breathing mode, 
constructed
from the terms at first order that involve $\bar{V}_0$ with the 
centrifugal terms from the kinetic energy
cancelling, while the other frequency, $\bar{\omega}_{0^{+}}$, the center of
mass mode, is constructed
from centrifugal terms at first order with the 
terms that involve $\bar{V}_0$
cancelling.

\smallskip

\subsection{The $[N-1]$ sector frequencies $\bar{\omega}_{1^{\pm}}$.}

The normal mode eigenvalues in the $[N-1,1]$ sector, $\bar{\omega}_{1^{+}}$ and
$\bar{\omega}_{1^{-}}$, are associated with angular and radial particle-hole
excitations.  These two frequencies are the closest frequencies to
the extremely low phonon frequency occupied in ultracold regimes
and thus play a role in
setting up an excitation gap for these systems.  
It is again enlightening to analyze the
relative contributions of the interaction potential, $\bar{V}_0$, versus the
kinetic energy for these $[N-1,1]$ sector frequencies.
 The radial particle-hole frequency should depend strongly
on the strength of $\bar{V}_0$, 
as a single particle
is excited from the ensemble in a radial direction. For the angular
particle-hole excitation frequency, I expect to see strong
 dependence on $\bar{V}_0$
drop out of the final simplified analytic expression and the centrifugal
terms in the effective potential contribute.
This is demonstrated in Appendix~\ref{app:N-1sectorfreq}.

\smallskip

\subsection{The $[N-2,2]$ sector: the $\bar{\omega}_2$ frequency.}

The normal mode frequency in the $[N-2,2]$ sector, $\bar{\omega}_2$, is
 associated with the phonon compressional mode which has an extremely 
small frequency and 
thus is the only normal mode occupied by a gas of fermions at ultracold
temperatures.
This is an angular mode, so I expect its frequency to be
relatively independent of  $\bar{V}_0$ and to
have a strong dependence on centrifugal terms. This is shown in Appendix~\ref{app:N-2sectorfreq}.

\section{The evolution of the excitation gaps from weakly interacting to unitarity}\label{sec:discussion}

I  discuss the emergence, growth and stability of excitation gaps
as the frequencies evolve as a function of the strength
of interparticle
interactions.

\subsection{$\bar{V}_0$ increases for a fixed ensemble size.}\label{subsec:V0increases}

I fix the system size, i.e. the value
of $N$, and let the interaction strength $\bar{V}_0$ increase. This analysis is
directly relevant to experiments which use a Feshbach resonance to tune
the interaction strength for a particular system. The value of
$\bar{V}_0$ is changed from essentially
zero, i.e. the case of independent, non-interacting particles 
trapped in a harmonic
potential, to the large interactions ($\bar{V}_0 = 1.0$) of the unitary regime.

\twocolumngrid

\begin{figure}
\centering
\begin{subfigure}[h]{0.51\textwidth}
\includegraphics[scale=0.65]{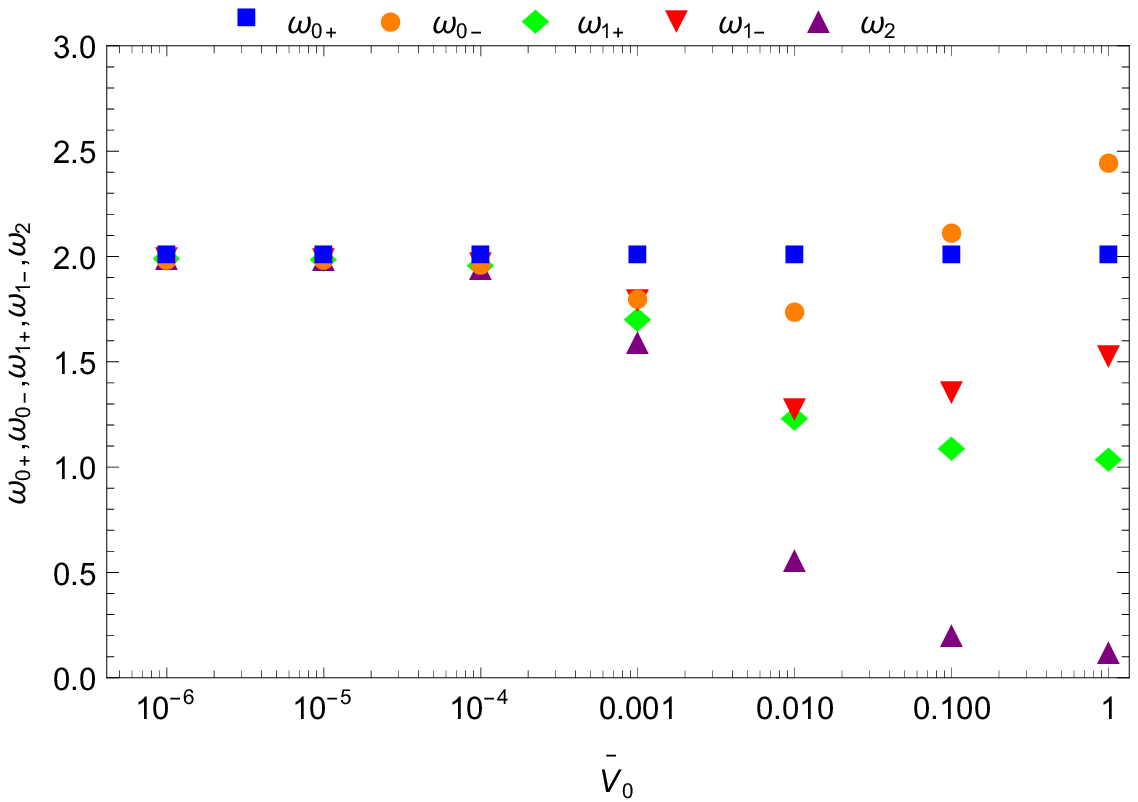}
\caption{a. $N=10^3$ fermions.}
\label{fig:trialthirteen}
\end{subfigure}
\begin{subfigure}[h]{0.51\textwidth}
\includegraphics[scale=0.65]{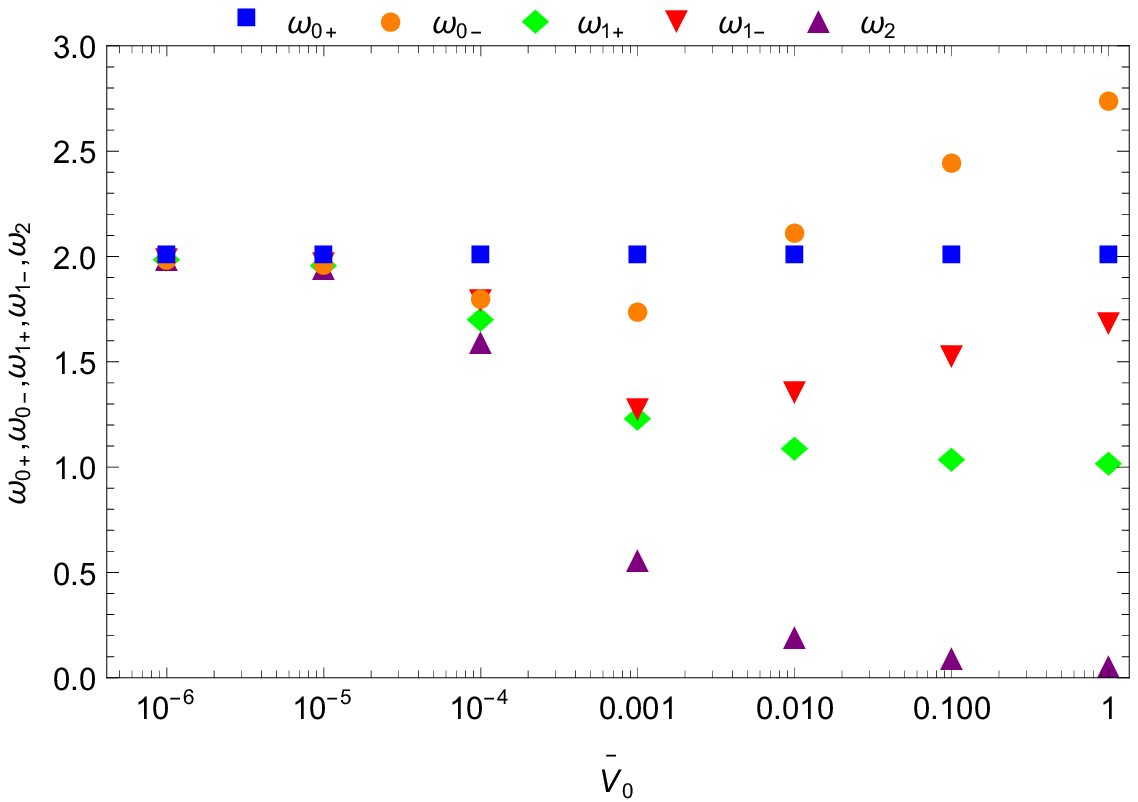}
\caption{b. $N=10^4$ fermions.}
\label{fig:trialfourteen}
\end{subfigure}
\begin{subfigure}[h]{0.51\textwidth}
\includegraphics[scale=0.65]{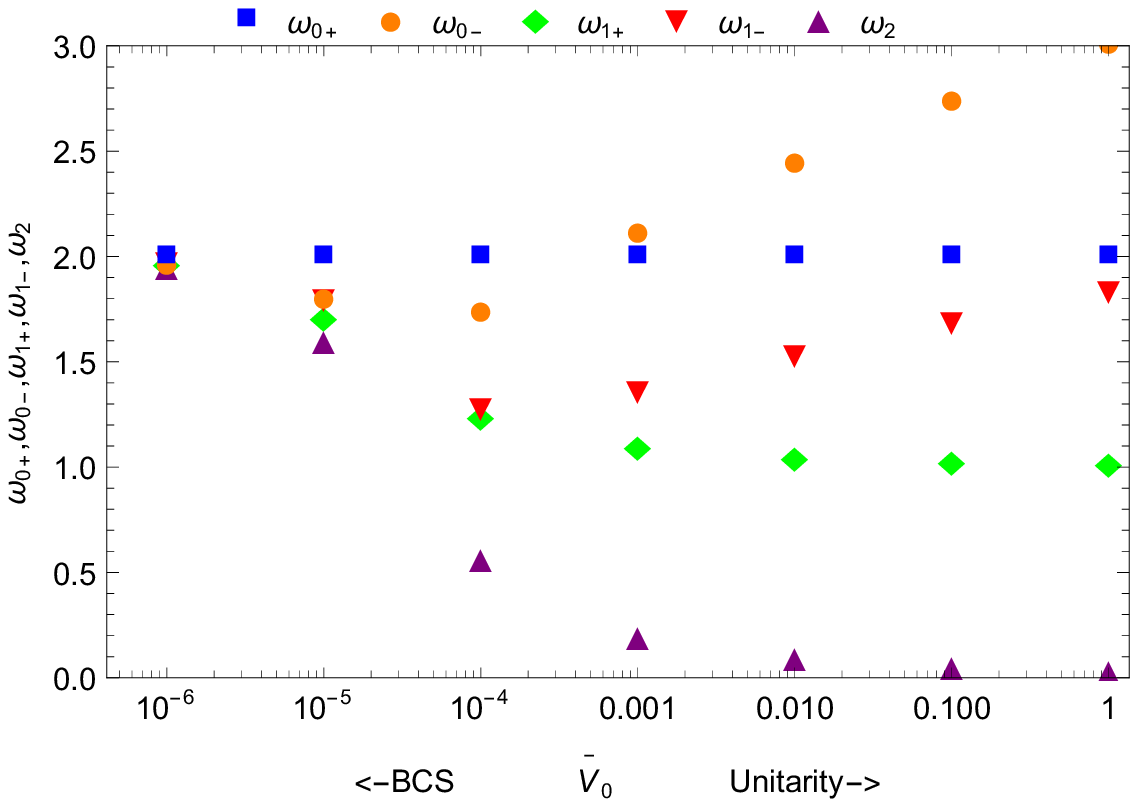}
\caption{c. $N=10^5$ fermions.}
\label{fig:trialfifteen}
\end{subfigure}
\setcounter{figure}{1}
\caption{Frequencies as a function of the strength of the interparticle
interaction, $\bar{V}_0$, from BCS to unitarity in units of the trap frequency.
Note the log scale on the x axis.}
\label{fig:eleven2}
\end{figure}

\twocolumngrid

\begin{figure}
\centering
\begin{subfigure}[h]{0.51\textwidth}
\includegraphics[scale=0.65]{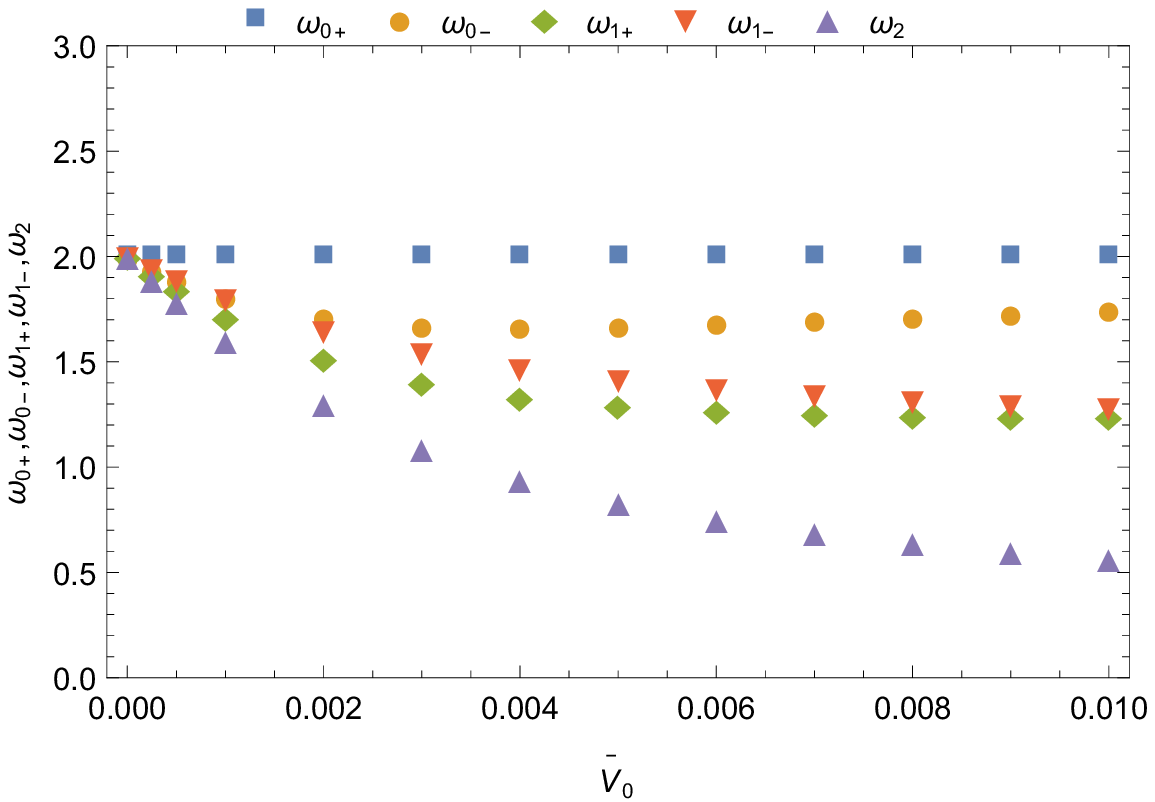}
\caption{a. $N=10^3$ fermions}
\label{fig:trialthirteenex}
\end{subfigure}
\begin{subfigure}[h]{0.51\textwidth}
\includegraphics[scale=0.65]{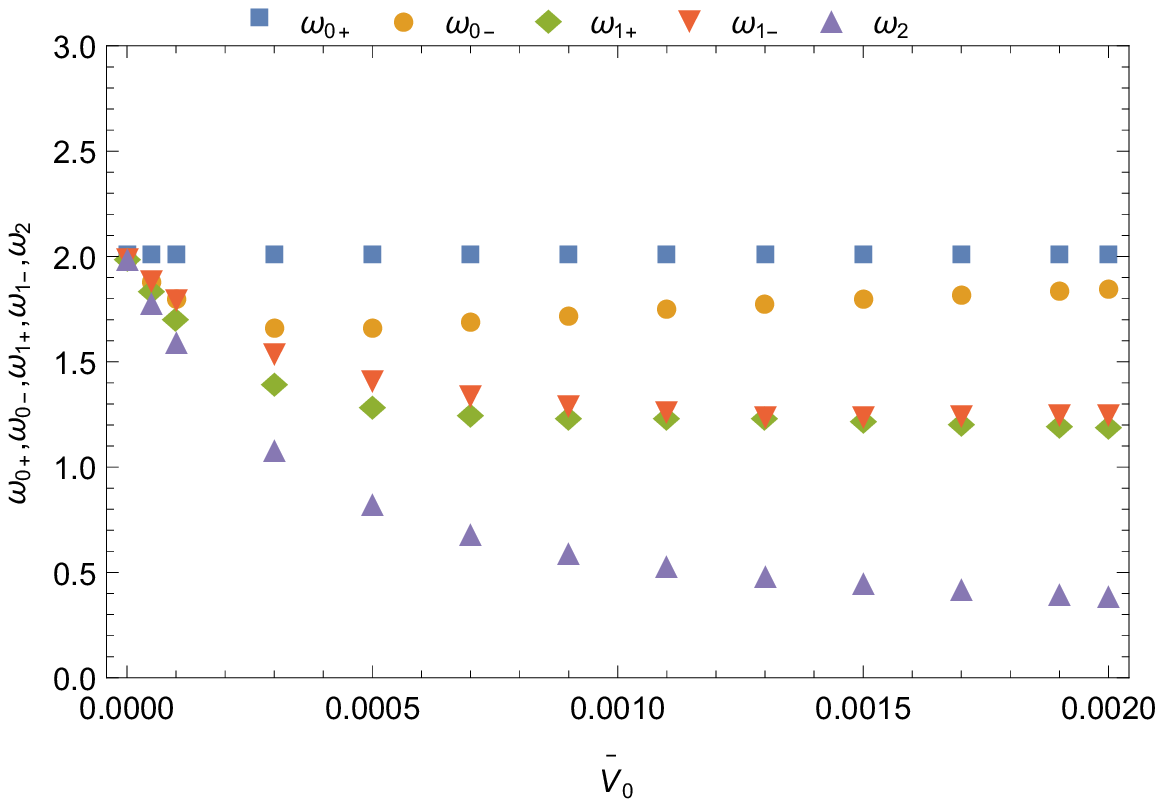}
\caption{b. $N=10^4$ fermions.}
\label{fig:trialfourteenex}
\end{subfigure}
\begin{subfigure}[h]{0.51\textwidth}
\includegraphics[scale=0.65]{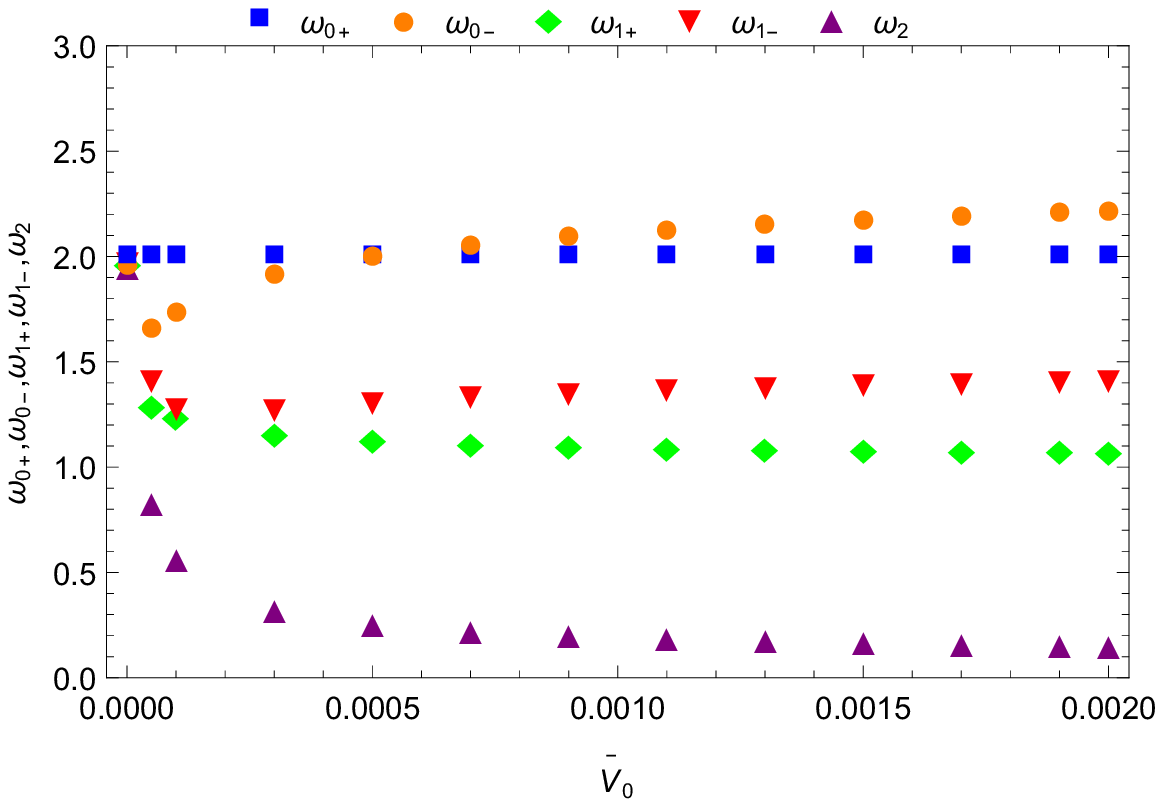}
\caption{c. $N=10^5$ fermions.}
\label{fig:trialfifteenex}
\end{subfigure}
\setcounter{figure}{2}
\caption{Expanded view of 
Figs.~(\ref{fig:trialthirteen})-(\ref{fig:trialfifteen}) 
near the independent particle region, i.e.
in the deep BCS regime, showing the rapid change in the frequencies
when interactions turn on. Note the small linear scale on the $x$ axis.}
\label{fig:thirteenex2}
\end{figure}

Figs.~(\ref{fig:trialthirteen})-(\ref{fig:trialfifteen}) show this effect   
for $N=10^3, \, 10^4$ and $10^5$ particles respectively
as $\bar{V}_0$ is tuned from the BCS regime to
the unitary regime.  The plots show that at the limit of zero 
interparticle interactions, the frequencies coalesce 
to the same value of $2\bar{\omega}_{ho}$ as expected and observed
in the laboratory\cite{bartenstein,thomas1}. 
(See an expanded view in 
Figs.~(\ref{fig:trialthirteenex})-(\ref{fig:trialfifteenex}) of the region 
near the independent
particle limit.) As interactions are slowly turned on, 
gaps rapidly emerge reaching values that stabilize for the angular
frequencies as unitarity is approached.

Unlike the angular frequencies that quickly converge to limiting values,
the radial frequencies continue to slowly increase as unitarity is
approached, suggesting that higher order terms are needed to converge the radial
frequencies. The angular
frequencies approach limits that are
integer multiples of the trap frequency: twice the
trap frequency for the center of mass angular frequency, 
$\bar{\omega}_{{0}^+} = 2\bar{\omega}_{ho}$; equal to the trap frequency for the single 
particle
angular excitation, $\bar{\omega}_{{1}^+} = \bar{\omega}_{ho}$; and  
orders of magnitude smaller than the trap frequency for the phonon mode, 
$\bar{\omega}_{2} = O(10^{-2})\bar{\omega}_{ho}$.  For $N \gg 1$, stable values 
for the angular
frequencies are reached quite quickly as $\bar{V}_0$ increases from the 
BCS regime. (See Figs.~(\ref{fig:trialthirteenex})-(\ref{fig:trialfifteenex}).)

As can be seen in all the above figures,  the largest gap forms 
between the extremely 
low frequency phonon mode which is the only mode occupied by ultracold gases 
and the next lowest frequency which is a particle-hole excitation, i.e. a 
single particle excitation. Both of these are angular frequencies which
reach stable limits, not changing as $\bar{V}_0$ increases.
This particular excitation gap is relevant to the emergence and 
sustainability of superfluidity
and is shown in Figs.~(\ref{fig:trialsixteen}) and (\ref{fig:trialseventeen}) 
for 
system sizes of $N=10^3$ and $N=10^4$ particles.
Note this gap emerges at lower 
 $\bar{V}_0$ i.e. for  weaker interactions for larger ensemble sizes. 
  
In the expanded view
of the region near the independent particle limit 
(Figs.~(\ref{fig:trialthirteenex})-(\ref{fig:trialfifteenex})) 
when the interactions
have just turned on, one notices two phenomena.  First the change in
the frequencies is quite rapid as soon as the interactions turn on 
(note the small scale on the x-axis), quickly
approaching values that will stabilize or change very slowly as the
interactions increase. Second, the frequencies separate more quickly for 
larger systems, i.e. as more and more particles are responding
 to a particular interaction strength.

\smallskip

{\it Thus, increasing the interaction between a fixed number of
 particles or increasing 
the number
of particles experiencing a fixed interaction has a similar effect in
separating the frequencies quickly.} (See Sec.~\ref{subsec:dynamics} for
a discussion of the microscopic dynamics underpinning these two approaches and 
Appendix~\ref{app:limits} for details of the analytic derivation of this 
effect.)

\twocolumngrid

\begin{figure}
\centering
\begin{subfigure}[h]{0.51\textwidth}
\includegraphics[scale=0.55]{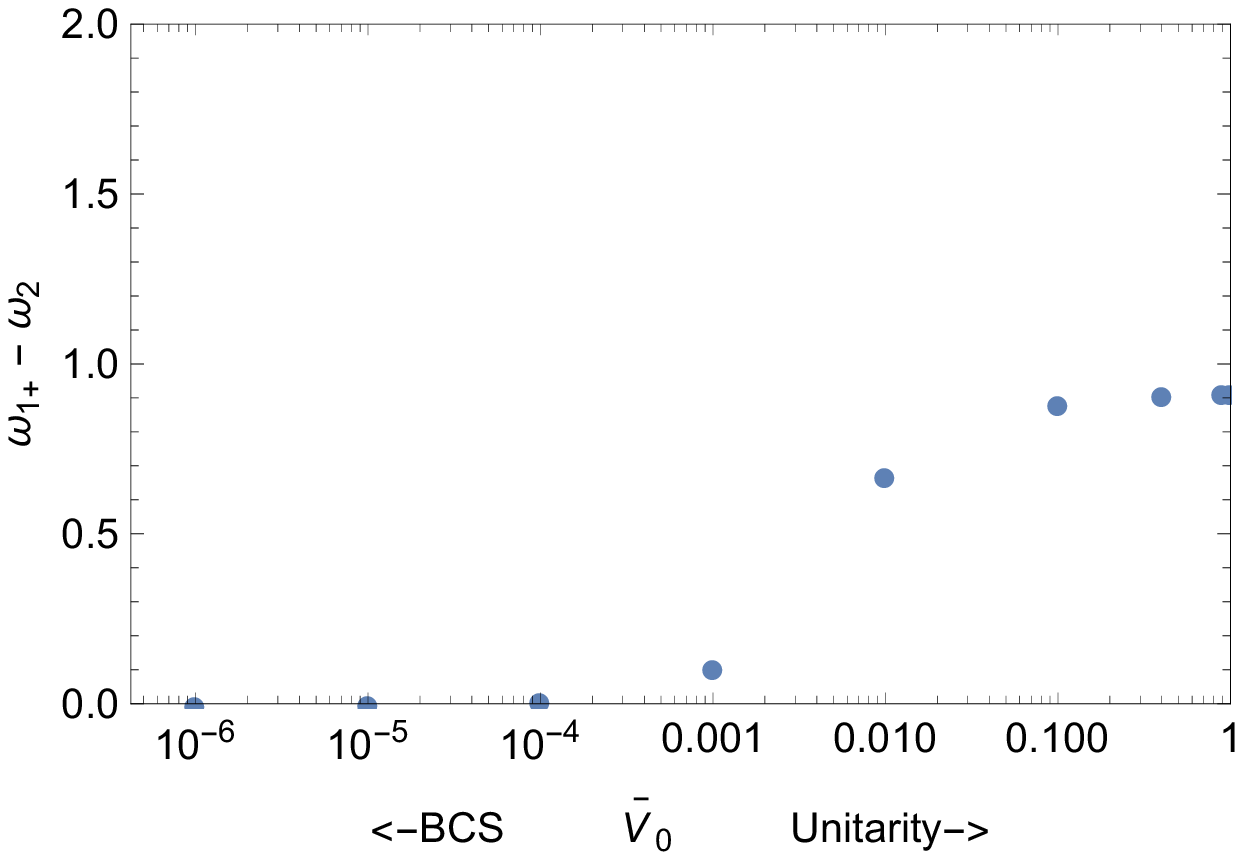}
\caption{a. $N=10^3$ fermions.}
\label{fig:trialsixteen}
\end{subfigure}
\begin{subfigure}[h]{0.51\textwidth}
\includegraphics[scale=0.55]{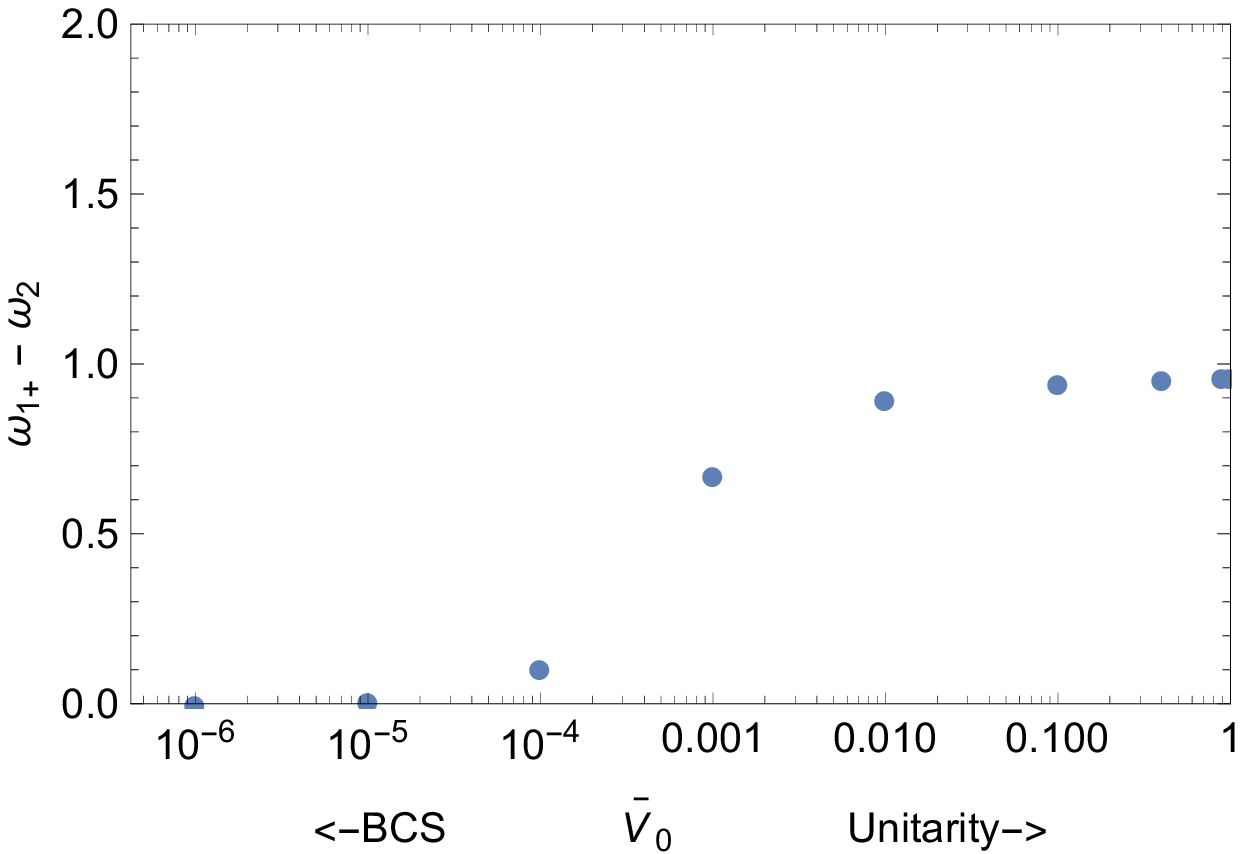}
\caption{b. $N=10^4$ fermions.}
\label{fig:trialseventeen}
\end{subfigure}
\setcounter{figure}{3}
\caption{Excitation gap in units of the trap frequency 
from the lowest two normal mode frequencies 
as a function of the strength of the interparticle
interaction, $V_0$, from BCS to unitarity.}
\label{fig:sixteen2}
\end{figure}

\subsection{Stable limits for the angular frequencies as a function of  $\bar{V}_0$.}\label{subsec:stablelimits}

The angular frequencies evolve from the independent particle limit to
stable limits at unitarity as  $\bar{V}_0$ increases. I will discuss
both these limits in this section, 
deriving them from the analytic 
expressions for the frequencies in Appendices~\ref{app:indeplimits} and 
\ref{app:limits} respectively. Then in Section~\ref{sec:micro}, I will 
take advantage of the analytic forms for both the normal mode frequencies
and the corresponding normal coordinates to understand 
 the microscopic dynamics underpinning the stability of these limits by
 tracking the evolution of behavior including the 
normal mode {\it motions} of individual particles as $\bar{V}_0$ increases.



 



\subparagraph{The independent particle limit: $\bar{V}_0 = 0$.}
Determining the values of the five frequencies is straightforward
in the limit of no interactions between the particles. Setting $\bar{V}_0$
equal to zero in the transcendental equations for  
$\gamma_{\infty}$ and $\bar{r}_{\infty}$ 
(See  Eqs.~(\ref{eq:minimum2})-(\ref{eq:gammaeq0bec})) results in values of 
$\gamma_{\infty}=0$ and $\bar{r}_{\infty}=1/\sqrt{2}$. Using these values
in the formulas for the frequencies, (See Appendix~\ref{app:indeplimits}.)
yields  a value of $2\bar{\omega}_{ho}$, an integer multiple of the trap
frequency, as expected\cite{thomas1,bartenstein,baranov} for each of the 
five frequencies since the only potential
affecting the particles is the harmonic trap. 
The individual fermions obey Fermi-Dirac statistics, but have no
interactions with the other fermions in the trap. Thus all 
five frequencies coalesce to the same
value. 
This can be clearly seen in 
Figs.~(\ref{fig:trialthirteen})-(\ref{fig:trialfifteen}) and
in the expanded view in 
Figs.~(\ref{fig:trialthirteenex})-(\ref{fig:trialfifteenex}).

\subparagraph{The unitary limit: $\bar{V}_0 =1.0$.}
As $\bar{V}_0$ is turned on and the particles begin to interact,
 the frequencies spread apart. The radial frequencies, $\bar{\omega}_{0^-}$ 
and $\bar{\omega}_{1^-}$ 
increase while the angular frequencies evolve to limits of
$\bar{\omega}_{{0}^+} = 2\bar{\omega}_{ho}$, $\bar{\omega}_{{1}^+} = 
\bar{\omega}_{ho}$, and 
$\bar{\omega}_{2} = O(10^{-2})\bar{\omega}_{ho}$. These limits for the angular
frequencies are stabilized at lower values of the interaction
strength for larger values of $N$ as previously discussed and as shown
in Figs.~(\ref{fig:trialthirteen})-(\ref{fig:trialfifteen}). 
In Figs.~(\ref{fig:trialthirteenU})-(\ref{fig:trialfourteenU}) the 
approach to the unitary regime 
on a linear scale shows the stability of the angular frequencies
and the gradual change in the radial frequencies. As will be demonstrated
in the next Section, the stable limits for the angular frequencies as
$\bar{V}_0$ increases signifies the vanishing of the interparticle
interactions for these angular motions. A derivation of these
limits from the analytic expressions for the angular frequencies 
is given in Appendix~\ref{app:limits}.

\smallskip

In the following section, the microscopic behavior that underpins this 
stability is analyzed using the analytic expressions 
for the angular frequencies and the motions 
 as analyzed in detail in Ref.~\cite{annphys}. 

\twocolumngrid

\begin{figure}
\centering
\begin{subfigure}[h]{0.51\textwidth}
\includegraphics[scale=0.7]{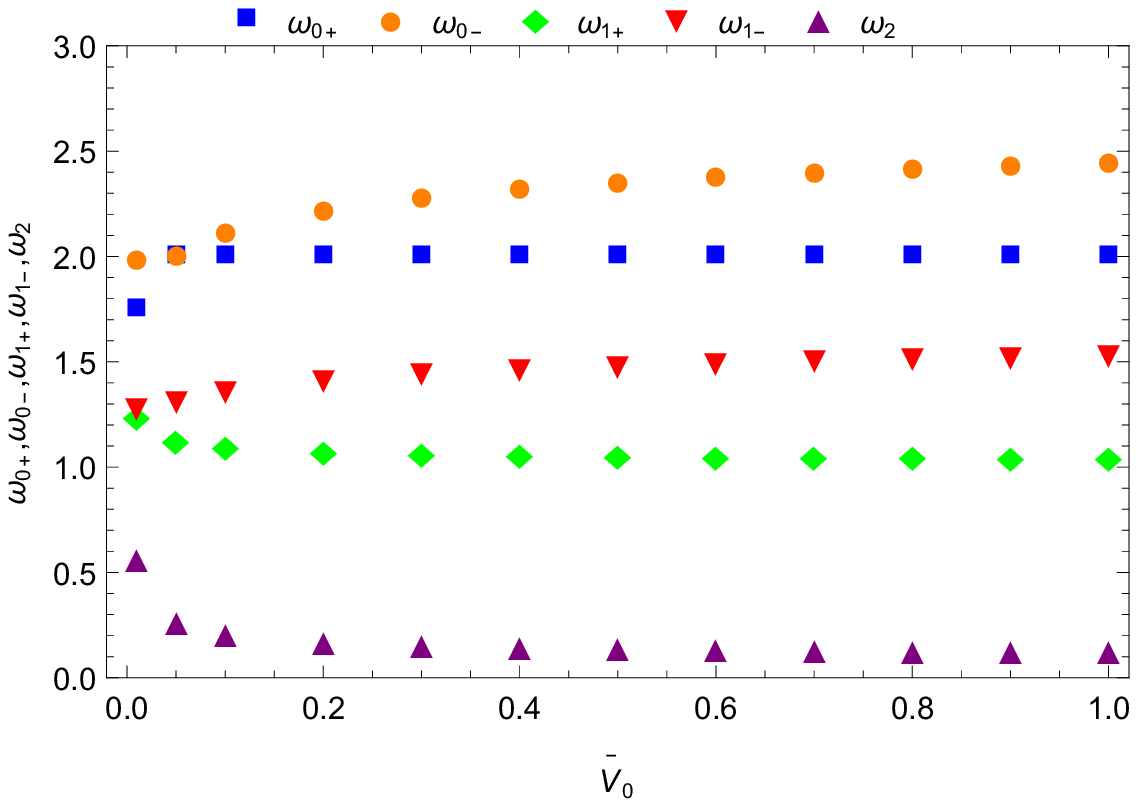}
\caption{a. $N=10^3$ fermions}
\label{fig:trialthirteenU}
\end{subfigure}
\begin{subfigure}[h]{0.51\textwidth}
\includegraphics[scale=0.7]{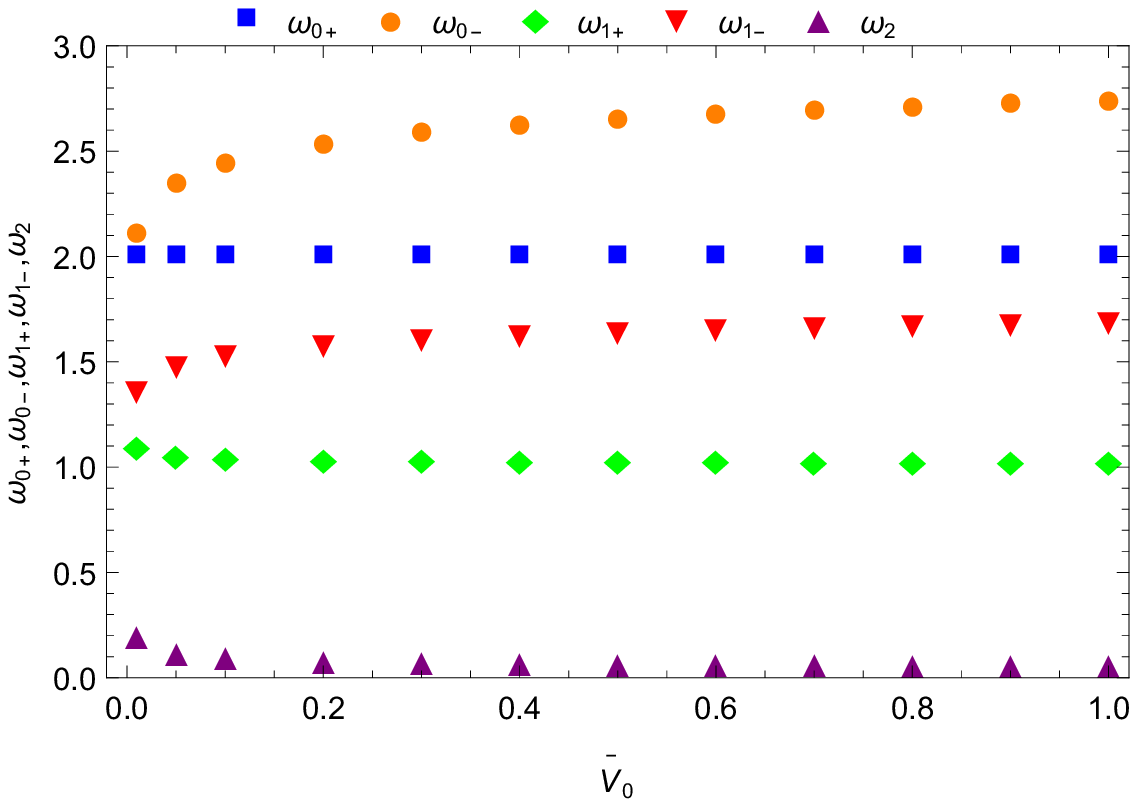}
\caption{b. $N=10^4$ fermions.}
\label{fig:trialfourteenU}
\end{subfigure}
\setcounter{figure}{4}
\caption{Expanded view of 
Figs.~(\ref{fig:trialthirteen}) and (\ref{fig:trialfourteen}) 
as unitarity is approached showing the stability of the angular frequencies
$\bar{\omega}_{{0}^+}$, $\bar{\omega}_{{1}^+}$ and $\bar{\omega}_2$,
and the gradual change of the radial frequencies
$\bar{\omega}_{{0}^-}$ and $\bar{\omega}_{{1}^-}$.}
\label{fig:thirteenUx2}
\end{figure}

\section{Understanding the microscopic dynamics that result in
stable limits for the angular frequencies.}\label{sec:micro}

The behavior of the angular frequencies as shown in 
Figs.~(\ref{fig:trialthirteen})-(\ref{fig:trialfifteen}) 
and discussed in the previous sections 
has revealed three interesting phenomena. 

\smallskip

\noindent 1) First, the angular frequencies are evolving
to stable limits independent of interactions as $\bar{V}_0$ increases 
while the radial
frequencies continue to slowly change. 

\smallskip

\noindent 2) Second, the limits for the angular
frequencies are {\it integer} multiples of the trap frequency. 

\smallskip

\noindent 3) Third, the gap between the frequencies emerges at weaker
interaction strengths for larger values of $N$.

\smallskip

I will now analyze the origins of these three interrelated phenomena 
for each of the three angular frequencies
%
by first looking at the 
analytic expressions for the frequencies 
in these limits; then by tracking the increase in 
correlation using the
variable $\gamma_{\infty}$; and finally by analyzing the 
corresponding {\it motion} of the 
associated normal mode
using the analysis in Ref.~\cite{annphys}.  The motion of the individual particles
offers an understanding of the
microscopic dynamics responsible for these phenomena, including the
emergence, growth and stability of the excitation gaps as $\bar{V}_0$ increases.

\subparagraph{Tracking the contributions of terms in the Hamiltonian
to the analytic expressions for the angular frequencies.} 
It is enlightening to analyze the evolution of 
the various terms in the Hamiltonian as they contribute
to the value of the frequencies as $\bar{V}_0$
increases. I will focus on terms in the
effective potential which is composed of three terms: the centrifugal term 
that originates in the kinetic
energy, the trap potential, and the interparticle potential; and focus on their
effect on the angular frequencies that are relevant to the emergence
of superfluid behavior in ultracold regimes. The trap
potential affects all the frequencies of course; the two radial frequencies
explicitly, and the angular frequencies implicitly through other variables.
If there are no interparticle interactions, all the frequencies would, 
of course, be integer multiples of the trap frequencies. 

The analysis in
 Appendices~\ref{app:Nsectorfreq}, \ref{app:N-1sectorfreq} and
\ref{app:N-2sectorfreq}
shows that  $\bar{V}_0$ contributes at first order to the 
radial frequencies (See Eq.~(\ref{eq:omega0mFG}) and (\ref{eq:omega1mFG})),
while cancelling out of the expressions for the angular frequencies at first order
which are dominated by the centrifugal potential terms.
The remaining explicit dependence on $\bar{V}_0$ ( in $F_g$)
for the angular frequencies
is small, 
damped by a factor of $1/N$. 
However, there are implicit dependences on $\bar{V}_0$ through the variables
 $\bar{r}_{\infty}$, $\gamma_{\infty}$,
$\tanh\Theta_{\infty}$ and $\sech^2\Theta_{\infty}$. 
Of these four variables, two of the variables, $\tanh\Theta_{\infty}$ and $\sech^2\Theta_{\infty}$, play a role
only in these damped terms. 

Of the two remaining variables, $\bar{r}_{\infty}$ and $\gamma_{\infty}$, 
the most interesting variable to study is
 $\gamma_{\infty}$, the angle cosine of each pair of particles in the
large dimension maximally symmetric configuration. This variable was identified
in early dimensional scaling work to signify the
existence of correlation between the particles\cite{herschbach,laingdensity,zhen}. 
The term, {\it correlation energy},
has been defined, for example, by
comparing the energies obtained in configuration interaction 
calculations with Hartree Fock/mean-field energies. 
The correlation energy reflects the change in energy as
the particles in the system move in a correlated way, thus minimizing their
interactions.   
These early studies compared dimensional scaling results to Hartree Fock 
mean field results and
noted that $\gamma_{\infty} = 0$ in the
Hartree Fock approximation which is an independent particle approximation. 
The non-zero values of $\gamma_{\infty}$ 
at zeroth order in the dimensional expansion thus indicated that
some correlation effects were being included even at lowest order underpinning
the excellent results obtained by this early work at 
low order\cite{loeser, kais1,kais2}.

\subparagraph{Tracking the magnitude of $\gamma_{\infty}$.} Tracking the
 magnitude
of $\gamma_{\infty}$, which is a negative quantity,
 as $\bar{V}_0$ increases and its effect on the different
terms in the expression for the frequencies has the potential to
 reveal insight into how the ensemble is adjusting {\it microscopically} to
the introduction of interactions between the particles, specifically how
the particles are rearranging 
 from ``independent'' motion to collective, 
coherent motion as correlation sets in. 
The Pauli principle is, of course, in effect as this transition occurs.  Its 
role is fundamental and will be addressed in depth in a subsequent study.
Each expression for the three
angular frequencies involves several terms that involve $\gamma_{\infty}$
from the $F$ and $G$ elements as seen in 
Eqs.~(\ref{eq:omega0pFG}), (\ref{eq:omega1pFG}) and (\ref{eq:omega2FGapp}) and 
Eqs.~(\ref{eq:lambda0pFG})-(\ref{eq:lambda2FG}). 
 These terms evolve as correlation increases and the magnitude
of $\gamma_{\infty}$ increases,  changing the relative
contributions of the kinetic energy, the trap and the interparticle
interaction to the frequency. I will analyze the response of
each angular frequency to these changes in $\gamma_{\infty}$ as correlation
increases below.

\subparagraph{Analyzing the microscopic motions of the angular normal modes
as unitarity is approached.}

What are the microscopic dynamics that are controlling this evolution to stable 
large gaps at unitarity? My analysis of the {\it motions} of the
normal modes in a recent study\cite{annphys} makes it possible to understand
the dynamics at a microscopic level as the ensemble rearranges its motion
from the independent particle case to correlated, collective behavior.
This motion is determined by fairly simple analytic expressions 
(See Eqs. (19)-(25) in Ref.~\cite{annphys}.) that involve 
an intricate balancing of Kronecker delta functions and Heaviside 
functions that give zero or unity depending on the integer 
indices that refer to specific particles. This accounting  keeps
 track of the interplay of all the particles, one-by-one and offers
a microscopic view of the dynamics leading to unitarity.

\subsection{The center of mass frequency 
$\bar{\omega}_{{0}^{+}}$.}\label{subsec:centerofmass}

The center of mass frequency 
is too large to be relevant for the emergence and support 
of superfluidity of ultracold gases. 
However
it is helpful to analyze its simple motion and 
well-known independence from interparticle interactions to
gain insight into the dynamics of the other angular normal modes.

\smallskip

\noindent{\it Independent of $\bar{V}_0$.} The center of mass frequency is 
independent of interparticle interactions.  
The system of particles
moves as a rigid body, all particles moving in lockstep with
identical motions. This frequency separates out as a constant value
in all the figures  at twice the trap frequency for all values of $N$ and
 $\bar{V}_0$. The analytic formula for $\bar{\omega}_{{0}^{+}}$ from the solution
of the first-order Hamiltonian is analyzed in
Appendix~\ref{app:Nsectorfreq} and reveals the cancellation of the terms
involving $\bar{V}_0$ to first order yielding an analytic expression
that is insensitive to changes in the interparticle interaction strength. 

\smallskip

\noindent{\it Independent of $N$.} The center of mass frequency is also constant as the system
size increases.  This is expected since a system that is independent
of interparticle interactions should behave like the independent particle
case with all particles responding only to the trap, not to other interactions.
Thus adding particles has no effect on the interactions felt by the other
particles and thus no effect on
the frequencies. (The fermions, of course, obey the Pauli principle which
affects the properties of the ensemble through Fermi-Dirac statistics but 
does not affect the frequencies.)

\smallskip

\noindent{\it Microscopic dynamics.} The individual particles in the center of mass normal
mode can be seen from my analysis of the normal mode motions in 
Ref.~\cite{annphys}, Sections 3.2 and 5.1, to be 
executing {\it identical} very small angular motions as a rigid body
with negligible change in their
radial {\it interparticle} distances.  Thus, these particles are not 
affected by the interparticle interaction resulting in a frequency that is an
{\it integer multiple} ($ = 2$) of the trap frequency since the trap 
is the only potential affecting the particles.

\smallskip

\noindent{\it Response of $\bar{\omega}_{{0}^{+}}$ to changes in 
$\gamma_{\infty}$.}
The center of mass
frequency remains at a fixed value of $\bar{\omega}_{{0}^+} = 2\bar{\omega}_{ho}$
 as the variables in the analytic
expression: 
\begin{eqnarray}
\bar{\omega}_{{0}^{+}} & \approx & \sqrt{\left[G_g+2(N-2)G_h\right]}   \\
&& \times \sqrt{\left[F_g+2(N-2)F_h+\frac{(N-2)(N-3)}{2}F_{\iota}\right]} 
\nonumber
\end{eqnarray}
change in response to system parameters. 
How does this expression remain fixed as its 
various terms are changing? 

Consider the independent particle limit where $\bar{V}_0 =0$,
$\gamma_{\infty} = 0$, $\bar{r}_{\infty}=1/\sqrt{2}$, and the
particles are affected only by the harmonic trap while 
obeying the Pauli principle.
Most of the terms in the
expression for  $\bar{\omega}_{{0}^{+}}$ are zero.  The only non-zero terms
are $G_g$ and $F_g$ (See Appendix~\ref{app:indeplimits}.) 
which both originate in the kinetic energy, depend implicitly 
on the trap potential,
and involve a dependence on just two particles, $i \,\mbox{and}\, j$, through 
the variable, $\gamma_{ij}$.  As $\bar{V}_0$ turns on, 
$\gamma_{\infty}$ takes on a small
nonzero value, signifying that weak correlations exist. This nonzero
value now means that all of the terms in $\bar{\omega}_{{0}^{+}}$ are nonzero
and $G_g$ and $F_g$ evolve to new values. 
Specifically $G_h, \,\, F_h\,\, \mbox{and}\,\, F_i$ acquire nonzero values 
and involve the dependence of the kinetic energy ($G_h$) and the
centrifugal potential ($F_h$ and $F_i$) on
$\gamma_{ij}\gamma_{jk}$ i.e involving three particles,
$i,j, \mbox{and}\, k$; and on $\gamma_{ij}\gamma_{kl}$  involving 
four particles, $i,j,k, \mbox{and}\, l$. These terms become significant in
determining the value of the frequency with factors of $2(N-2)$ for $F_h$
and $(N-2)(N-3)/2$ for $F_i$ as
interparticle correlations increase. Since the value of 
$\bar{\omega}_{{0}^{+}}$  remains fixed at 2,
the magnitudes of  $G_g$ and $F_g$ adjust as correlations spread out
throughout the ensemble. Thus, the emergence of interparticle interactions 
starts an intricate readjustment of the ensemble as the particles 
governed by the evolving Hamiltonian respond to the other $N-1$ particles. 

\subsection{The angular particle-hole excitation frequency $\bar{\omega}_{{1}^{+}}$.}\label{subsec:angparticlehole}
Now consider the angular single-particle excitation frequency 
$\bar{\omega}_{{1}^{+}}$
which can also be described as a particle-hole angular excitation.
\smallskip

\noindent{\it Independent of $\bar{V}_0$.} 
This frequency reaches a constant value 
equal to the trap frequency as $\bar{V}_0$
increases reflecting the vanishing of interparticle interactions. 
The analytic formula for $\bar{\omega}_{{1}^{+}}$ analyzed in
Appendix~\ref{app:N-1sectorfreq} reveals the cancellation of the terms
involving $\bar{V}_0$ to first order.  Thus this frequency is expected to
become constant as the interaction changes if higher order terms are small.

\smallskip

\noindent{\it Independent of $N$.} 
It also becomes constant as the system
size increases since the frequency is insensitive to all interparticle 
interactions so additional particles have no effect.

\noindent{\it Microscopic dynamics.} This behavior can be understood from a 
microscopic view of the motions of the
particles. In this case, the motion of
the corresponding normal mode is made up of one particle creating a
``large'' angular displacement with the other particles, while the remaining 
interparticle angles make a small adjustment. The first group has $N-1$ 
interparticle angles, while the second group of $(N-1)(N-2)/2$ angles 
quickly becomes the overwhelming majority of the ensemble with 
displacements that are smaller by a factor of $(N-2)/2$. 
(See Sections 3.4 and 5.1 in Ref.~\cite{annphys})
These two opposing and
unequal motions invoke some radial interactions from slight changes
in the interparticle distances and thus this mode does have
some response to the interaction $\bar{V}_0$.  However for values of $N$
typical of laboratory ensembles ($10^4-10^6$), the percentage of 
particles moving in lockstep by a smaller
and smaller angular amount becomes so dominant that the radial 
contribution is insignificant. Thus the harmonic trap is the dominant
effect determining this frequency, analogous to
the center of mass frequency.
This explains the value of the frequency at an {\it integer multiple} 
($ = 1$) of the trap
frequency and its independence from changes in $\bar{V}_0$ and/or $N$.

\smallskip

\noindent{\it Response of $\bar{\omega}_{{1}^{+}}$ to changes in 
$\gamma_{\infty}$.}
 When $\bar{V}_0 = 0$ and the
particles are moving independently, $\gamma_{\infty}=0$ and 
$\bar{r}_{\infty}=1/\sqrt{2}$, only $G_g$ and $F_g$ in the
expression for  $\bar{\omega}_{{1}^{+}}$ are nonzero with values determined
by the harmonic trap yielding $2\bar{\omega}_{ho}$, a integer multiple
of the trap frequency for $\bar{\omega}_{{1}^{+}}$. 
(See Appendix~\ref{app:indeplimits}.)
\begin{eqnarray}
\bar{\omega}_{{1}^{+}} & \approx & \sqrt{\left[G_g+(N-4)G_h\right]}  \\
&&\times \sqrt{\left[F_g+(N-4)F_h-(N-3)F_{\iota}\right]} \nonumber \\
&\rightarrow & 2\bar{\omega}_{ho} \,\,\,\,\,\,(G_g=4, F_g=1, G_h=F_h=F_i=0) \nonumber
\end{eqnarray}
 As interparticle interactions are introduced, 
$\gamma_{\infty}$ is no longer zero so $G_h, \,\, F_h\,\, \mbox{and}\,\, F_i$ 
contribute as
interparticle correlations increase. Smaller factors for $F_h$ of $(N-4)$
and a negative factor of $-(N-3)$ for $F_i$  result in a smaller 
 value of $\bar{\omega}_{{1}^{+}}$ which evolves
 to $\bar{\omega}_{ho}$ from its value of 2$\bar{\omega}_{ho}$ at the independent
particle limit. The magnitudes of  $G_g$ and 
$F_g$ adjust as the changing value of $\gamma_{\infty}$ signifies
longer-range correlations bringing in new
contributions involving three and four particles. 

\subsection{The phonon frequency $\bar{\omega}_2$.}\label{subsec:phonon}
Finally consider the angular phonon compressional frequency $\bar{\omega}_2$.
 This frequency reaches a constant value
that is two or three orders of magnitude
smaller than the trap frequency as $\bar{V}_0$ increases.

\smallskip
 
\noindent{\it Independent of $\bar{V}_0$.}
The analytic formula for $\bar{\omega}_2$:
\begin{equation}
\bar{\omega}_2 = \sqrt{\left[G_g-2G_h\right]
 \left[F_g-2F_h+F_{\iota}\right]} \label{eq:omega2FG}  \nonumber
\end{equation}
analyzed in
Appendix~\ref{app:N-2sectorfreq} reveals the insignificance of the terms
involving $\bar{V}_0$ to first order.  Thus this frequency is expected to
become constant as the interaction increases.

\smallskip

\noindent{\it Independent of $N$.}
It also becomes constant as the system
size increases since adding particles has no effect on this frequency
which is independent of the interparticle interaction to first order.

\smallskip

\noindent{\it Microscopic dynamics.}
In this third case, the motion of
the corresponding normal mode is made up of three groups of interparticle
angles  involving particles that move
with different angular motions and amounts:
 a single dominant interparticle angle which has the largest
angular displacement;   $2(N-2)$ nearest neighbor
angles which move 
with an opposing angular displacement
that is smaller than that of the dominant angle by a factor of $(N-2)$; and
a third group which quickly becomes the dominant group of particles 
involving $(N-2)(N-3)/2$ angles which have a displacement that is a factor 
of $(N-2)(N-3)/2$
smaller than the dominant angle. (See Sections 3.5 and 5.1-5.1.3
 in Ref.~\cite{annphys}.) 
These groups move in opposing directions
with unequal displacements and thus experience some radial interparticle 
interactions.  However as $N$
increases, the percentage of particles moving in lockstep in the third group
by a smaller
and smaller angular amount becomes so dominant that the radial 
contribution due to the movement of the other two groups is negligible.
(See Section 5.1.3 in Ref.~\cite{annphys}.) 
This results in the value of the frequency at an {\it integer multiple 
($\approx 0$)} of the trap
frequency and independence from changes in $\bar{V}_0$ and/or $N$.

\smallskip

\noindent{\it Response of $\bar{\omega}_2$ to changes in $\gamma_{\infty}$.}
 When $\bar{V}_0 = 0$ and the
particles are moving independently affected only by the trap potential, $\gamma_{\infty}=0$ and 
$\bar{r}_{\infty}=1/\sqrt{2}$.  The non-zero terms, $G_g$ and $F_g$,
depend on $\gamma_{ij}$. (See Appendix~\ref{app:indeplimits}.) 
As $\bar{V}_0$ turns on, 
$\gamma_{\infty}$ is
nonzero so all terms in 
$\bar{\omega}_2$ are now nonzero and involve a dependence on
$\gamma_{ij}\gamma_{jk}$ and $\gamma_{ij}\gamma_{kl}$. Since the value of 
$\bar{\omega}_2$ 
evolves to very small values from $2\bar{\omega}_{ho}$,
the magnitudes of  $G_g$ and $F_g$ must adjust as longer-range 
correlations 
throughout the ensemble reflect the realignment of the 
particles into collective motion.

\subsection{A discussion of the microscopic dynamics}\label{subsec:microdiscussion}
The dynamics that drive the angular frequencies to integer multiples of the
trap frequency at unitarity are responsible for both the large 
excitation gap between the lowest two normal modes and the 
independence of the ensemble
from the microscopic interaction details consistent with the expected 
universal behavior.
There are, in fact, two distinct dynamical effects
 that can produce this behavior.
The discussion of the microscopic dynamics in the above paragraphs 
assumes large values of $N$ typical of
experiments with ultracold Fermi gases to understand these stable limits
as the interparticle interactions vanish. However, increasing $\bar{V}_0$
 can also
result in the angular frequencies approaching integer multiples of the 
trap frequency for fixed values of $N$. These two effects can be
seen in the figures in 
 Section~\ref{sec:discussion} which show $\bar{V}_0$ increasing for
several fixed ensemble sizes. This complimentary behavior as either
$N$ or $\bar{V}_0$ increases was previously noted at the end of 
Section~\ref{subsec:V0increases} and an analytic derivation of these 
two effects
is given in Appendix~\ref{app:limits}.  
I discuss both these behaviors in more detail
below. 

\subparagraph{Two distinct microscopic dynamics.}\label{subsec:dynamics}
As discussed above, when $N$ increases to large values, 
the percentage of particles that have very small angular movements, i.e.
$\gamma_{ij}\ll 1\, $ in the $[N-1,1]$ and $[N-2,2]$ angular modes 
become the overwhelming majority of
particles in the ensemble. 
(The center of mass mode, of course, has all the particles moving
in lockstep with amounts that are small when $N$ is large.)
The angular motion i.e, the magnitude of $\gamma_{ij}$, for this majority
of particles 
becomes smaller and smaller as $N$ increases. (See Section 5.1 in
Ref~\cite{annphys}.)  
Since {\it purely} angular motions produce no change in the radial 
distances from the center of the trap, i.e.
$\bar{r}_i$ and $\bar{r}_j$ are constant, 
this motion yields negligible changes in the interparticle distances,
$\bar{r}_{ij}={\sqrt{{\bar{r}_i}^2+
{\bar{r}_j}^2-2\bar{r}_i\bar{r}_j\gamma_{ij}}}$ when $\gamma_{ij}$ is tiny.
%
\smallskip

Now consider letting $\bar{V}_0$ increase for fixed system size.
The correlation between
the particles increases as tracked by the parameter $\gamma_{\infty}$.
This happens quite rapidly as $\bar{V}_0$ increases from zero reflecting
the rearrangement of the particles into correlated angular motion as collective
behavior sets in. This dynamic is relevant to experiments using
Feshbach resonances to tune interactions to the 
large values of the unitary regime.
As correlations spread throughout the ensemble into this rigid angular 
motion, the
interparticle interactions become negligible for this fixed value of $N$ and
the system is independent of the details of the microscopic interactions.

\smallskip

\subparagraph{Microscopic dynamics of unitarity.} The lowest two normal 
modes in the unitary regime have frequencies that
set up an excitation gap that is stable and independent of the microscopic
details of the interaction. The spectrum of the $[N-1,1]$ angular mode 
has evenly spaced levels at {\it every integer multiple of the trap frequency,
 identical to the spectrum
of the non-interacting regime of independent particles.}  
The strong interactions
of the unitary regime result in synchronized, correlated behavior
that paradoxically have minimal interparticle interactions.

\smallskip

The unitarity limit is defined as having no interaction length scale due
to strong interactions that are much shorter range than the interparticle distance,
leaving the oscillator length and the interatomic distance as the only
relevant length scales. The interatomic distance is the defining
characteristic of an angular normal mode which has all (center of mass mode)
or the overwhelming majority 
(angular particle-hole excitation and phonon modes) of the particles
moving as a rigid body with collisionless motion.
The gas is expected to show a universal thermodynamic
behavior at zero temperature, independent of any microscopic details of the 
underlying interactions.



\section{Summary and Conclusions} \label{sec:SumConc}

In this study, I have looked in detail at the analytic frequencies for $N$
identical particles as a function of the interparticle interaction strength
as it is tuned from weakly interacting regimes to the strong interactions of the
unitary regime. The frequencies were obtained previously from
the normal mode solutions to the SPT
first-order equation in inverse dimensionality for a
system of confined, interacting, identical particles.  
These $N$-body normal modes were 
determined analytically as a function of various system parameters 
and used to construct wave functions and
density profiles for systems of 
identical bosons \cite{paperI,JMPpaper,laingdensity}
and later energies\cite{prl} and thermodynamic quantities 
for fermions\cite{partition,emergence}.

The current investigation is motivated by a recent study of the
evolution of the N-body analytic 
normal mode {\it coordinates} as $N$ increases from
 few-body systems that have good molecular
equivalents to the expected behavior of many-body ensembles\cite{annphys}.
A specific Hamiltonian, that of the unitary regime, was investigated and
 two phenomena were noted
that could sustain the emergence and stability of superfluid behavior.
In this paper, I have extended the study of these two phenomena
to a range of  
interaction strengths from BCS to unitarity.

In particular, I have investigated closely the behavior of the lowest two
angular frequencies that are relevant to the emergence of excitation gaps
that could support superfluidity as the interparticle strength is
increased from BCS to unitarity.  I used both the analytic expressions 
for the {\it frequencies} which allow the different
contributions from Hamiltonian terms to be assessed and 
the simple analytic expressions for the normal mode 
{\it motions}\cite{annphys} to gain insight into the microscopic dynamics 
underpinning this evolution.

\smallskip

\subparagraph{Summary.} In summary, my analysis 
has resulted in a number of observations 
 that may prove
useful in understanding the emergence, growth and stability
 of excitation gaps as well as offering a possible explanation of 
the microscopic dynamics responsible for
universal behavior at unitarity. I list them below:

\smallskip

\noindent 1) The analytic expressions for the normal mode frequencies produce 
behavior that supports the emergence of excitation gaps consistent with
 the known behavior of ultracold Fermi gases
 in the laboratory tuned using Feshbach 
resonances from the weakly interacting BCS regime with small gaps 
to the large gaps of
the strongly interacting unitary regime.

\smallskip

\noindent 2) The normal modes evolve to almost purely radial or purely
angular character as $N$ increases, with very little mixing of the symmetry 
coordinates,  over the entire transition from BCS to unitarity. 
This confirms that 
the frequencies can be labelled as {\it radial}
or {\it angular} and affects the stability since the symmetry coordinates
are analytic solutions of an underlying, approximate Hamiltonian.

\smallskip

\noindent 3) As $\bar{V}_0$ increases from zero at the independent particle 
limit, these first-order analytic frequencies 
rapidly separate. As unitarity is approached the angular frequencies stabilize
 while the
radial frequencies continue to slowly change. This suggests that higher order 
terms may be necessary
to converge the radial frequencies.

\smallskip

\noindent 4) The change in the frequencies 
emerges at weaker interaction strengths as the ensemble grows.  
Thus, increasing the number
of particles experiencing a fixed interaction or 
increasing the interaction between a fixed number of
 particles  has a similar effect in
separating the frequencies quickly.

\smallskip

\noindent 5) The largest gap forms between the extremely 
low frequency angular phonon mode, which is the only mode occupied by 
ultracold gases, 
and the next lowest frequency which is an angular
 particle-hole excitation, i.e. a 
single-particle excitation. This particular excitation gap is relevant to the 
emergence and 
sustainability of superfluidity in ultracold systems. 

\smallskip

\noindent 6) The limits for the three angular
frequencies are integer multiples of the trap frequency, reflecting the
{\it interaction independence} of these frequencies. 

\smallskip

\noindent 7) The lowest two normal mode frequencies 
relevant to ultracold gases provide a spectrum 
of evenly spaced levels at integer multiples of the trap frequency at
unitarity, identical
to the spectrum of the non-interacting regime of independent particles.  This
spectrum is, of course, independent of any microscopic details of the underlying
interactions consistent with the dynamics expected for the unitary regime.
Thus, the strong interactions of the unitary regime result in strong,
long-range
correlated behavior that paradoxically has minimal interparticle
interactions.

\smallskip

\noindent 8) Two distinct dynamical effects were found that 
can drive the angular frequencies to integer multiples of the trap
frequency at unitarity. First, when $N$ increases,  
the angular phonon and single-particle
excitation modes that are involved in creating an excitation gap
for ultracold particles now have an overwhelming percentage of particles
 moving with small, purely angular motions 
that have a negligible
response to interparticle interactions. This results in angular frequencies 
at  integer multiples of the trap frequency 
since the trap is the only potential affecting the particles.
Second, as $\bar{V}_0$ increases for fixed $N$,
correlations increase and become long range as tracked by the parameter,
$\gamma_{\infty}$.  The motion evolves into rigid angular motion
as interparticle interactions vanish
 yielding frequencies at integer multiples of the trap
frequency.

\smallskip


\subparagraph{Conclusions.}

This analysis of the normal mode frequencies 
yields consistent, physically intuitive behavior that has been observed
in the laboratory. 
The microscopic dynamic underlying this behavior 
is based on normal mode motions and thus is different than the 
accepted view that
the relevant particles in a superfluid form loosely bound pairs
that decrease in size as a Feshbach resonance is tuned to strong 
interactions. 

Normal modes have an infinite spectrum of evenly-spaced excited states. 
At unitarity, the
spectrum involved in an excitation gap for ultracold fermions
consists of integer steps of the trap frequency identical to the 
spectrum of the non-interacting independent particle limit. This
behavior supports dynamics at unitarity that is independent of 
interparticle interactions. Despite having the same spectrum, the
dynamics of independent fermions in a trap 
are quite different from the dynamics of fermions at unitarity whose behavior
reflects the strong
interactions that have been encapsulated into 
normal mode motions.
A full understanding of how this spectrum affects the dynamics at unitarity 
requires an understanding  of 
the role of the Pauli principle which is the subject of a future study.


If higher order effects are small, the normal coordinates whose frequencies
and mixing coefficients depend on the interparticle interactions are, in fact, 
beyond-mean-field ${\it analytic}$ solutions to a many-body Hamiltonian.
The frequencies and the
motions of the normal modes evolve in sync with each other, both
responding to the same microscopic dynamics. 
These analytic forms for the frequencies and coordinates allow the 
details of the terms in the Hamiltonian that are driving the
dynamics to be revealed in a particularly transparent way.

Specifically, I looked at the change
in the parameter $\gamma_{\infty}$ whose magnitude increases as $\bar{V}_0$
increases signalling an increase in the strength and long-range character of  
correlation as terms
involving three and four particles begin to contribute to the values
of the frequencies.

The dynamics revealed by this study are based on a {\it exact} solution of the
the first-order equation of SPT perturbation theory. 
If higher order terms
are significant in a particular regime along this transition, the dynamics 
could change. In particular, the radial frequencies (which are not
involved in providing excitation gaps for ultracold gases) do not show stable
limits as unitarity is approached which suggests that higher order terms
are needed for these frequencies. First order SPT results have been tested only
in the unitary regime, i.e. for strong interactions,
yielding  ground state energies comparable to benchmark Monte Carlo
results\cite{prl} and excellent agreement with experiment
for thermodynamic quantities\cite{emergence}.  The weakly interacting
regime has so far been unexplored using this formalism. 
This approach also does not offer a mechanism for the pairing in real space 
that occurs beyond unitarity
as the ensemble transitions to the BEC regime.


Normal mode functions 
provide simple, coherent macroscopic wave functions with phase coherence 
over the entire system. 
The dynamics of a normal mode description of the BCS to unitarity
transition with its many-body pairing offers an interesting 
alternative to the models 
relying on two-body pairing
mechanisms to achieve superfluidity. This approach also offers 
a possible microscopic understanding of the 
universal behavior at unitarity which could be applicable to other 
strongly correlated 
superfluids in diverse systems.

\section{Acknowledgments}
I would like to thank the National Science Foundation for financial support
under Grant No. PHY-1607544 and Grant No. PHY-2011384.

\bigskip

\appendix
\renewcommand{\theequation}{A\arabic{equation}}
\setcounter{equation}{0}

\section{The mixing coefficients for the $[N]$ and $[N-1,1]$
sectors.}
\label{app:mixing}

The mixing coefficients for the $[N]$ sector are:

\begin{widetext}
\begin{eqnarray}
\mbox{cos}\theta^{[N]}_+ &=& \frac{\sqrt{2} \sqrt{N-1}(c+(N/2-1)d)}
{\sqrt{2(N-1)(c+(N/2-1)d)^2
+(-a-(N-1)b+{\lambda}_{[N]}^+)^2}}\label{eq:cos0p} \\
\mbox{sin}\theta^{[N]}_+ &=& \frac{-a-(N-1)b+{\lambda}_{[N]}^+}
{\sqrt{2(N-1)(c+(N/2-1)d)^2+(-a-(N-1)b
+{\lambda}_{[N]}^+)^2}} \label{eq:sins0p}\\
\mbox{cos}\theta^{[N]}_- &=& \frac{\sqrt{2} \sqrt{N-1}(c+(N/2-1)d)}
{\sqrt{2(N-1)(c+(N/2-1)d)^2
+(-a-(N-1)b+{\lambda}_{[N]}^-)^2}}\label{eq:cos0m} \\
\mbox{sin}\theta^{[N]}_- &=& \frac{-a-(N-1)b+{\lambda}_{[N]}^-}
{\sqrt{2(N-1)(c+(N/2-1)d)^2+(-a-(N-1)b
+{\lambda}_{[N]}^-)^2}}\,, \label{eq:sin0m}
\end{eqnarray}
\end{widetext}

while the coefficients in the 
$[N-1,1]$ sector are:

\begin{widetext}
\begin{eqnarray}
\mbox{cos}\theta^{[N-1,1]}_+ &=&  \frac{\sqrt{N-2}(c-d)}
{\sqrt{(N-2)(c-d)^2 + (-a+b+{\lambda}_{[N-1,1]}^+)^2}}\label{eq:cos1p} \\
\mbox{sin}\theta^{[N-1,1]}_+ &=& \frac{-a+b+{\lambda}_{[N-1,1]}^+}
{\sqrt{(N-2)(c-d)^2 + (-a+b+{\lambda}_{[N-1,1]}^+)^2}}\label{eq:sin1p} \\
\mbox{cos}\theta^{[N-1,1]}_- &=& \frac{\sqrt{N-2}(c-d)}
{\sqrt{(N-2)(c-d)^2 + (-a+b+{\lambda}_{[N-1,1]}^-)^2}}\label{eq:cos1m} \\
\mbox{sin}\theta^{[N-1,1]}_- &=&  \frac{-a+b+{\lambda}_{[N-1,1]}^-}
{\sqrt{(N-2)(c-d)^2 + (-a+b+{\lambda}_{[N-1,1]}^-)^2}}.\label{eq:sin1m} 
\end{eqnarray}
\end{widetext}

\noindent where ${\lambda}_{[N]}^\pm$ and ${\lambda}_{[N-1,1]}^\pm$ are
 given by Eqs.~(\ref{eq:omega_p})-(\ref{eq:lam1defs}) in
Section~\ref{sec:analyticfrequencies}.

The above equations have some explicit $N$ dependence (but no
$\bar{V}_0$ dependence) that is 
due to the symmetry present in the first-order Hamiltonian. 
The quantities $a,b,c,d,e,f,g,h,i$ in the expressions for the mixing
coefficients and the eigenvalues,
 ${\lambda}_{[N]}^\pm$ and ${\lambda}_{[N-1,1]}^\pm$, 
are defined in Appendix~\ref{app:FG}
(See also Eq. (42) in Ref~\cite{FGpaper}.)
in terms of the $F$ and $G$ elements and have explicit $N$ and 
$\bar{V}_0$ dependence as well
as $N$ and $\bar{V}_0$ dependence from the $F$ and $G$ elements from a 
particular Hamiltonian.

Thus there are three layers of analytic expressions
that can bring in $N$ and/or $\bar{V}_0$ dependence: the expressions for mixing coefficients
 in Eqs.~(\ref{eq:cos0p})-(\ref{eq:sin1m}) above, 
the expressions for $a,b,c,d,e,f,g,h,\iota,  \lambda_{[N]}^{\pm}
\mbox{ and } \lambda_{[N-1,1]}^{\pm}$
and the expressions for the $F$ and $G$ elements for a specific Hamiltonian.

\renewcommand{\theequation}{B\arabic{equation}}
\setcounter{equation}{0}

\section{The FG matrix elements.}\label{app:FG}

The constants used in the expressions for the mixing coefficients and the
frequencies are defined below:
\begin{eqnarray}\label{eq:GFsym}
a&=&{G}_{a}{F}_{a}   \nonumber \\
b&=&{G}_{a}{F}_{b} \nonumber\\
c&=&{G}_{g}{F}_{e} +(N-2){G}_{h}({F}_{e} + {F}_{f}) \nonumber\\
d&=&{G}_{g}{F}_{f} + 2{G}_{h}({F}_{e} + (N-3){F}_{f})  \nonumber\\
e&=&{G}_{a}{F}_{e}  \\
f&=&{G}_{a}{F}_{f}  \nonumber \\
g&=&{G}_{g}{F}_{g}+2(N-2){G}_{h}{F}_{h} \nonumber\\
h&=&{G}_{g}{F}_{h}+{G}_{h}{F}_{g}+(N-2){G}_{h}{F}_{h}+(N-3){G}_{h}{F}_{\iota} \nonumber\\
\iota&=&{G}_{g}{F}_{\iota}+4{G}_{h}{F}_{h}+2(N-4){G}_{h}{F}_{\iota}.
\nonumber
\end{eqnarray}

The non-zero elements
of the ${\bf G}$ matrix are:
\begin{eqnarray}
\label{eq:G_HS}
{G}_{a}&=& G_{\bar{r}_i\bar{r}_i}=1  \nonumber \\
{G}_{g}&=& G_{\gamma_{ij}\gamma_{ij}} = 2\frac{1-{\gamma_{\infty}}^2}{{\bar{r}_{\infty}}^2} \nonumber \\
&=& 4[1+(N-1)\gamma_{\infty}](1+\gamma_{\infty})(1-\gamma_{\infty}) \nonumber \\
{G}_{h}&=& G_{\gamma_{ij}\gamma_{jk}} = \frac{\gamma_{\infty}(1-\gamma_{\infty})}{{\bar{r}_{\infty}}^2} \nonumber \\
&=& 2[1+(N-1)\gamma_{\infty}]\gamma_{\infty}(1-\gamma_{\infty}) \,, \nonumber
\end{eqnarray}
where the matrix elements have been evaluated at the infinite-$D$
symmetric minimum.

\noindent Likewise the non-zero ${\bf F}$ matrix elements are:
\begin{eqnarray}
\label{eq:F_HS} {F}_{a}&=&\left(\frac{\partial^2 \bar{V}_\text{eff}}
{\partial\bar{r}_i^2}\right)\Biggr|_{\infty} \nonumber \\
 &=& 1+\frac{3}{4 \bar{r}_{\infty}^4}
\frac{1+(N-2)\gamma_{\infty}}{(1-\gamma_{\infty})(1+(N-1)\gamma_{\infty})} \\
&& +\frac{\bar{V}_{o} \bar{c}_{o}}{2}(N-1) \mbox{sech}^2\Theta_{\infty}
\Biggl[\bar{c}_{o}(1-\gamma_{\infty}) \tanh\Theta_{\infty} - \nonumber\\
&\; \;& - \frac{1+\gamma_{\infty}} {2\bar{r}_{\infty}\sqrt{1-\gamma_{\infty}}}
\Biggr] \nonumber \\
{F}_{b}&=&\left(\frac{\partial^2 \bar{V}_\text{eff}}{\partial\bar{r}_i\partial\bar{r}_j}\right)\Biggr|_{\infty} \nonumber \\
&=& \frac{\bar{V}_{o} \bar{c}_{o}}{2}\mbox{sech}^2\Theta_{\infty}
\Biggl[\bar{c}_{o}(1-\gamma_{\infty}) \tanh\Theta_{\infty} +
\frac{1+\gamma_{\infty}} {2\bar{r}_{\infty}\sqrt{1-\gamma_{\infty}}}
\Biggr] \nonumber \\
{F}_{e}&=&\left(\frac{\partial^2 \bar{V}_\text{eff}}{\partial\bar{r}_i\partial\gamma_{ij}}\right)\Biggr|_{\infty} \nonumber \\
&=& -\frac{\gamma_{\infty}}{2 \bar{r}_{\infty}^3}
\frac{1+(N-2)\gamma_{\infty}}{(1-\gamma_{\infty})^2(1+(N-1)\gamma_{\infty})^2} \\
&& +\frac{\bar{V}_{o} \bar{c}_{o}}{2} \mbox{sech}^2\Theta_{\infty}
\Biggl[-\bar{c}_{o}\bar{r}_{\infty} \tanh\Theta_{\infty}
+ \frac{1} {2 \sqrt{1-\gamma_{\infty}}}
\Biggr] \nonumber \\
{F}_{f}&=&\left(\frac{\partial^2 \bar{V}_\text{eff}}{\partial\bar{r}_i\partial\gamma_{jk}}\right)\Biggr|_{\infty}\nonumber \\
&=&\frac{{\gamma_{\infty}}^2}
{2\bar{r}_{\infty}^3(1-\gamma_{\infty})^2(1+(N-1)\gamma_{\infty})^{2}} \\
{F}_{g}&=&\left(\frac{\partial^2\bar{V}_\text{eff}}{\partial\gamma_{ij}^2}
\right)\Biggr|_{\infty}
\nonumber \\
&=&\frac{1}{2\bar{r}_{\infty}^2(1-\gamma_{\infty})^3(1+(N-1)\gamma_{\infty})^{3}} \nonumber \\
&& \times\Bigl(1+3(N-2)\gamma_{\infty} + (13-11N+3N^2){\gamma_{\infty}}^2 \nonumber \\
 &\; \;& + (N-2)(4-3N+N^2){\gamma_{\infty}}^3\Bigr) + \frac{\bar{V}_{o} \bar{c}_{o}}{2}\mbox{sech}^2\Theta_{\infty} \nonumber \\
&& \times \Biggl[\frac{\bar{c}_{o}\bar{r}^2_{\infty}}{1-\gamma_{\infty}}
\tanh\Theta_{\infty} + \frac{\bar{r}_{\infty}}{2(1-\gamma_{\infty})^{3/2}}
\Biggr] \nonumber \\
{F}_{h}&=&\left(\frac{\partial^2 \bar{V}_\text{eff}}{\partial\gamma_{ij}\partial\gamma_{jk}}\right)\Biggr|_{\infty}\nonumber \\
&=&\frac{-\gamma_{\infty}}
{4\bar{r}_{\infty}^2(1-\gamma_{\infty})^3(1+(N-1)\gamma_{\infty})^{3}} \\
&& \times \left[3+(5N-14)\gamma_{\infty} + (11-9N+2N^2){\gamma_{\infty}}^2
\right],\nonumber\\
F_{\iota}&=&\left(\frac{\partial^2 \bar{V}_\text{eff}}{\partial\gamma_{ij}\partial\gamma_{kl}}\right)\Biggr|_{\infty}\nonumber \\
&=& \frac{\gamma_{\infty}^2
(2+(N-2)\gamma_{\infty})}{\bar{r}_{\infty}^2(1-\gamma_{\infty})^3
(1+(N-1)\gamma_{\infty})^3} 
\end{eqnarray}

Inspection of the formulas for $F$ elements easily reveals the explicit 
dependence of their terms on the confining (trap) potential,
$\bar{V}_{\mathtt{conf}}$ , the centrifugal potential $\bar{U}$ and/or the 
interparticle interaction potential, $\bar{V}_0$. All the terms also have
some {\it implicit} dependence on all the terms in $\bar{V}_{\mathtt{eff}}$ 
through the
variables, $\bar{r}_{\infty}$ and $\gamma_{\infty}$. I list the {\it explicit}
contributions below:

\begin{eqnarray}
\label{eq:F_HSex} 
{F}_{a} & \Leftrightarrow & \,\bar{V}_{\mathtt{conf}},\, \bar{U}\, \mbox{and}\, \bar{V}_0 (strong) \nonumber \\
{F}_{b} & \Leftrightarrow & \,\bar{V}_0 (weak) \nonumber \\
{F}_{e} & \Leftrightarrow & \,\bar{U} \, \mbox{and}\, \bar{V}_0 (weak) \nonumber \\
{F}_{f} & \Leftrightarrow & \,\bar{U} \nonumber \\
{F}_{g} & \Leftrightarrow & \,\bar{U}\, \mbox{and}\, \bar{V}_0 (weak) \nonumber \\
{F}_{h} & \Leftrightarrow & \,\bar{U} \nonumber \\
F_{\iota} & \Leftrightarrow & \,\bar{U} \nonumber
\end{eqnarray}

Using the fact that the three nonzero $G$ elements and the 
centrifugal ``potential'' terms originate in kinetic energy terms of
the Hamiltonian, it is possible to classify the $a,b,c,d,e,f,g,h,\iota$
terms. Since they all contain one of the three $G$ elements, they all
have a contribution from the kinetic energy. Below I list the {\it explicit} 
contributions in terms of the potentials.

\begin{eqnarray}\label{eq:GFsymex}
a&=&{G}_{a}{F}_{a} \,\Leftrightarrow \,\bar{V}_{\mathtt{conf}},\,  \bar{U},\, \mbox{and} \,\bar{V}_0 (strong) \nonumber  \\
b&=&{G}_{a}{F}_{b}  \,\Leftrightarrow \, \bar{V}_0 (weak) \nonumber\\
c&=&{G}_{g}{F}_{e} +(N-2){G}_{h}({F}_{e} + {F}_{f}) \,\Leftrightarrow \,\bar{U}\, \mbox{and} \,\bar{V}_0 (weak) \nonumber \\
d&=&{G}_{g}{F}_{f} + 2{G}_{h}({F}_{e} + (N-3){F}_{f}) \,\Leftrightarrow \,\bar{U}\, \mbox{and} \,\bar{V}_0 (weak) \nonumber \\
e&=&{G}_{a}{F}_{e} \,\Leftrightarrow \,\bar{U}\, \mbox{and} \,\bar{V}_0 (weak) \nonumber \\
f&=&{G}_{a}{F}_{f}  \,\Leftrightarrow \,\bar{U}  \nonumber \\
g&=&{G}_{g}{F}_{g}+2(N-2){G}_{h}{F}_{h} \,\Leftrightarrow \,\bar{U} \, \mbox{and} \,\bar{V}_0 (weak) \nonumber \\
h&=&{G}_{g}{F}_{h}+{G}_{h}{F}_{g}+(N-2){G}_{h}{F}_{h}+(N-3){G}_{h}{F}_{\iota} 
\nonumber \\
&& \,\Leftrightarrow \,\bar{U}\, \mbox{and} \,\bar{V}_0 (weak) \nonumber\\
\iota&=&{G}_{g}{F}_{\iota}+4{G}_{h}{F}_{h}+2(N-4){G}_{h}{F}_{\iota} \,\Leftrightarrow \, \bar{U} \nonumber
\end{eqnarray}
\renewcommand{\theequation}{C\arabic{equation}}
\setcounter{equation}{0}
\section{The analysis of the $[N]$ sector frequencies.}
\label{app:Nsectorfreq}

The frequency, ${\bar{\omega}}_{{0}^{\pm}}$, associated with the root 
$\lambda_{{0}^{\pm}}$ of
multiplicity $1$ is given by:

\begin{equation}\label{eq:omega0pm}
\bar{\omega}_{{0}^{\pm}}=\sqrt{\eta_0 \mp
\sqrt{{\eta_0}^2-\Delta_0}},
\end{equation}

\begin{equation}\label{eq:lam0defsapp}
\begin{split}
\eta_0 &= \frac{1}{2}\Biggl[a+(N-1)b+g+2(N-2)h \\
& \,\,\,\,\,\,\,\,\,\,\,\,\,\,\,\,\,\,\,\,\,\,\,\,\,\,\,\,\,\,\,\,\, +\frac{(N-2)(N-3)}{2}\iota \Biggr] \\ 
\Delta_0 &= (a+(N-1)b)\left[g+2(N-2)h+\frac{(N-2)(N-3)}{2}\iota\right]  \\
&\,\,\,\,\,\,\,\,\,\,\,\,\,\,\,\,\,\,\,\,\,\,\,\,\,\,\,-\frac{N-1}{2}(2c+(N-2)d)(2e+(N-2)f).
\end{split}
\end{equation}
\begin{equation}\label{eq:etadelta0app}
\begin{split}
\eta_0^2 - \Delta_0 &= \frac{1}{4}\Biggl[a+(N-1)b-(g+2(N-2)h \\
& + \frac{(N-2)(N-3)}{2} \iota) \Biggr]^2 \\
& + \frac{(N-1)}{2}\left(2c+(N-2)d\right)\left(2e+(N-2)f\right)
\end{split}
\end{equation}
\noindent Defining:
\begin{eqnarray}
A &=& a+(N-1)b \nonumber \\
T &=& (g+2(N-2)h+\frac{(N-2)(N-3)}{2} \iota) \nonumber
\end{eqnarray}
\noindent and substituting the definitions of $a,b,c,d,e,f,g,h$ and $\iota$ 
in terms of $F$ and
$G$ elements:
\begin{eqnarray}
A &=& G_aF_a+(N-1)G_aF_b \nonumber \\
T &=& \left[G_g+2(N-2)G_h\right] \nonumber \\
&& \times \left[F_g+2(N-2)F_h+\frac{(N-2)(N-3)}{2}F_{\iota}\right] \nonumber \\
 &=& T_1 T_2 \nonumber \\
2c+(N-2)d &=& \left[G_g+2(N-2)G_h\right] 
 \left[2F_e+(N-2)F_f\right] \nonumber \\
&=& T_1 T_3 \nonumber \\
2e+(N-2)f &=&  \left[2F_e+(N-2)F_f\right] \nonumber \\
&=& T_3 \nonumber 
\end{eqnarray}
So:
\begin{eqnarray}
\bar{\omega}_{{0}^{\pm}}&=&\sqrt{\eta_0 \mp \sqrt{{\eta_0}^2-\Delta_0}}  \nonumber \\
&=&\sqrt{\frac{1}{2}\left[A+T\right] \mp \sqrt{\frac{1}{4}\left[A-T\right]^2 
+\frac{(N-1)}{2}T_1T_3^2}} \nonumber
\end{eqnarray}
\noindent Thus,
\begin{eqnarray}\label{eq:omegapm}
\omega_{{0}^{-}} &=&\sqrt{\frac{1}{2}\left[A + T\right] 
+ \frac{1}{2}\left[A - T\right](1+x)^{1/2}}  \nonumber \\
\omega_{{0}^{+}} &=& \sqrt{\frac{1}{2}\left[A + T\right] 
- \frac{1}{2}\left[A - T\right](1+x)^{1/2}}  \nonumber \\
\end{eqnarray} 
\noindent where 
\begin{equation} 
x = \frac{\frac{(N-1)}{2} T_1T_3^2}{\frac{1}{4}(A-T)^2} \nonumber
\end{equation}
\noindent is small.
Thus the two frequencies $\omega_{{0}^{\pm}}$ in the $[N]$ sector 
split into a frequency,
$\omega_{{0}^{-}}$, which has a leading term (under the square root) $A$ with 
a strong dependence on the interparticle
interaction potential, $\bar{V}_0$, 
and a frequency $\lambda_{{0}^{+}} $ 
 in which  $A$ cancels out leaving a leading term $T$ that depends on
the centrifugal terms. The powers of $x$ bring in
higher order terms.
\begin{eqnarray}
\omega_{{0}^{-}} & \approx & \sqrt{A}=\sqrt{a+(N-1)b}  \nonumber \\
& = & \sqrt{G_aF_a+(N-1)G_aF_b} \label{eq:omega0mFG} \\
\omega_{{0}^{+}} & \approx & \sqrt{T} =\sqrt{g+2(N-2)h+\frac{(N-2)(N-3)}{2} \iota} \nonumber \\
& = & \sqrt{\left[G_g+2(N-2)G_h\right]}  \label{eq:omega0pFG} \\
&& \times \sqrt{\left[F_g+2(N-2)F_h+\frac{(N-2)(N-3)}{2}F_{\iota}\right]} 
\nonumber
\end{eqnarray}
\renewcommand{\theequation}{D\arabic{equation}}
\setcounter{equation}{0}
\section{The analysis of the $[N-1,1]$ sector frequencies.}
\label{app:N-1sectorfreq}
The frequency, ${\bar{\omega}}_{{1}^{\pm}}$, associated with the root 
$\lambda_{{1}^{\pm}}$ of
\begin{equation}
\bar{\omega}_{{1}^{\pm}}=\sqrt{\eta_1 \mp
\sqrt{{\eta_1}^2-\Delta_1}},
\end{equation}
\begin{equation}\label{eq:lam1defsapp}
\begin{split}
\eta_1 &= \frac{1}{2}\Biggl[a-b+g+(N-4)h \\
& \,\,\,\,\,\,\,\,\,\,\,\,\,\,\,\,\,\,\,\,\,\,\,\,\,\,\,\,\,\,\,\,\, 
+(N-3)\iota \Biggr] \\ 
\Delta_1 &= (a-b)\left[g+(N-4)h-(N-3)\iota\right]  \\
&\,\,\,\,\,\,\,\,\,\,\,\,\,\,\,\,\,\,\,\,\,\,\,\,\,\,\,-(N-2)(c-d)(e-f),
\end{split}
\end{equation}
\begin{equation}
\begin{split}
\eta_1^2 - \Delta_1 & = \frac{1}{4}\bigl[a-b-(g+(N-4)h \\
& - (N-3)\iota) \bigr]^2 \\
& + (N-2)\left(c-d\right)\left(e-f\right)
\end{split}
\end{equation}
\noindent Defining:
\begin{eqnarray}
B &=& a-b \nonumber \\
R &=& (g+(N-4)h-(N-3)\iota) \nonumber 
\end{eqnarray}
\noindent and substituting the definitions of $a,b,c,d,e,f,g,h$ and $\iota$ 
in terms of $F$ and
$G$ elements:
\begin{eqnarray}
B &=& G_aF_a - G_aF_b \nonumber \\
R &=& \left[G_g+(N-4)G_h\right] \nonumber \\
&& \times \left[F_g+(N-4)F_h-(N-3)F_{\iota}\right] \nonumber \\
 &=& R_1 R_2 \nonumber \\
c-d &=& \left[G_g+(N-4)G_h\right] 
\times \left[F_e-F_f\right] \nonumber \\
&=& R_1 R_3 \nonumber \\
e-f &=&  \left[F_e-F_f\right] \nonumber \\
&=& R_3 \nonumber 
\end{eqnarray}
So:
\begin{equation}
\begin{split}
\bar{\omega}_{{1}^{\pm}}&=\sqrt{\eta_1 \mp \sqrt{{\eta_1}^2-\Delta_1}}  \nonumber \\
&=\sqrt{\frac{1}{2}\left[B+R\right] \mp \sqrt{\frac{1}{4}\left[B-R\right]^2 
+(N-2)R_1R_3^2}} \nonumber
\end{split}
\end{equation}
Regrouping the expressions for $\eta_1$ and $\eta_1^2-\Delta_1$ using these
factors yields:
\begin{eqnarray} 
\omega_{{1}^{-}} &=& \sqrt{\frac{1}{2}\left[B + R\right] 
+ \frac{1}{2}\left[B - R\right](1+x')^{1/2}}  \nonumber \\
\omega_{{1}^{+}} &=& \sqrt{\frac{1}{2}\left[B + R\right] 
- \frac{1}{2}\left[B - R\right](1+x')^{1/2}} \nonumber
\end{eqnarray}
\noindent where 
\begin{equation} 
x' = \frac{(N-2) R_1R_3^2}{\frac{1}{4}\left(B-R\right)^2} \nonumber
\end{equation}
\noindent is small.
Similar to the $[N]$ sector, the two frequencies $\omega_{{1}^{\pm}}$ 
in the $[N-1,1]$ sector 
split into a frequency,
$\omega_{{1}^{-}}$, which has a leading term  $B$ with 
a strong dependence on the interparticle
interaction potential, $\bar{V}_0$, and a frequency $\omega_{{1}^{+}} $ 
 in which  $B$ cancels out leaving a leading term $R$ that depends on
the centrifugal terms. The powers of $x'$ bring in
higher order terms.
\begin{eqnarray}
\omega_{{1}^{-}} & \approx &\sqrt{B} = \sqrt{a-b} \nonumber \\
&=& \sqrt{G_aF_a-G_aF_b} \label{eq:omega1mFG} \\
\omega_{{1}^{+}} & \approx &  \sqrt{R} = \sqrt{g+(N-4)h-(N-3)\iota} \nonumber \\ 
&=&\sqrt{\left[G_g+(N-4)G_h\right]} \label{eq:omega1pFG} \\
&&\times \sqrt{\left[F_g+(N-4)F_h-(N-3)F_{\iota}\right]} \nonumber 
\end{eqnarray}
\renewcommand{\theequation}{E\arabic{equation}}
\setcounter{equation}{0}
\section{The analysis of the $[N-2,2]$ sector frequency.}
\label{app:N-2sectorfreq}
The frequency, ${\bar{\omega}}_2$, associated with the root $\lambda_2$ of
multiplicity $N(N-3)/2$ is given by:
\begin{equation}\label{eq:omega2app}
\bar{\omega}_2=\sqrt{g-2h+\iota}.
\end{equation}
\noindent Substituting the definitions of $g,h$ and $\iota$ 
in terms of $F$ and
$G$ elements, the terms in the expression for $\bar{\omega}_2$ 
 can be factored as:
\begin{eqnarray}
\bar{\omega}_2 &=& \sqrt{\left[G_g-2G_h\right]
 \left[F_g-2F_h+F_{\iota}\right]} \label{eq:omega2FGapp} 
\end{eqnarray}
Note that the only term, $F_g$, that has explicit dependence on $\bar{V}_0$
has only a weak dependence, while $F_h$ and $F_{\iota}$ contain contributions 
from the centrifugal potential. As $\bar{V}_0$
approaches unitarity, this frequency decreases
becoming a tiny fraction of the trap frequency.
\renewcommand{\theequation}{F\arabic{equation}}
\setcounter{equation}{0}
\section{Limits for the frequencies at the independent particle
limit.}\label{app:indeplimits}
The formulas for the five frequencies from 
Section~\ref{sec:analyticfrequencies} are summarized below:
\begin{eqnarray}\label{eq:omegasapp}
\bar{\omega}_{{0}^{\pm}} &=& \sqrt{\eta_0 \mp \sqrt{{\eta_0}^2-\Delta_0}} \nonumber \\
\bar{\omega}_{{1}^{\pm}} &=& \sqrt{\eta_1 \mp \sqrt{{\eta_1}^2-\Delta_1}} \nonumber \\
\bar{\omega}_2 &=& \sqrt{g-2h+\iota}. 
\end{eqnarray}
\noindent where the variables $\eta_0, \Delta_0, \eta_1$ and $\Delta_1$ are defined in Eqs.~(\ref{eq:lam0defsapp}) and (\ref{eq:lam1defsapp}),
%
%
%
in terms of the quantities $a,b,c,d,e,f,g,h, \mbox{ and } \iota$ given
in Appendix~\ref{app:FG}  in terms of the $F$ and $G$ elements of the
first-order Hamiltonian.

At the independent particle limit, also referred to as the 
non-interacting limit, 
$\bar{V}_0 =0$, $\gamma_{\infty}=0$ and $\bar{r}_{\infty}=1/\sqrt{2}$.
Substituting these values into the $F$ 
and $G$ elements readily gives:
\begin{eqnarray}
\label{eq:F_HSapp} 
{F}_{a} & = & \,4 \,\,\,\,\,\,\,\,\,\,\,\,\,\,\,\, {G}_{a} =  1 \nonumber \\
{F}_{b} & = & \,0 \,\,\,\,\,\,\,\,\,\,\,\,\,\,\,\, {G}_{g} =  4 \nonumber \\
{F}_{e} & = & \,0 \,\,\,\,\,\,\,\,\,\,\,\,\,\,\,\, {G}_{h} =  0 \nonumber \\
{F}_{f} & = & \,0 \nonumber \\
{F}_{g} & = & \,1 \nonumber \\
{F}_{h} & = & \,0 \nonumber \\
F_{\iota} & = & \,0 \nonumber
\end{eqnarray}
%
%
\noindent Substituting these values into the expressions for $a,b,c,d,e,f,g,h,
\iota$ yields:
\begin{eqnarray}\label{eq:GFsymapp}
a&=&{G}_{a}{F}_{a} =  4  \nonumber \\
b&=&{G}_{a}{F}_{b} = 0 \nonumber\\
c&=&{G}_{g}{F}_{e} +(N-2){G}_{h}({F}_{e} + {F}_{f}) = 0 \nonumber\\
d&=&{G}_{g}{F}_{f} + 2{G}_{h}({F}_{e} + (N-3){F}_{f}) = 0 \nonumber\\
e&=&{G}_{a}{F}_{e}  = 0 \nonumber \\
f&=&{G}_{a}{F}_{f} = 0 \nonumber \\
g&=&{G}_{g}{F}_{g}+2(N-2){G}_{h}{F}_{h} = 4 \nonumber \\
h&=&{G}_{g}{F}_{h}+{G}_{h}{F}_{g}+(N-2){G}_{h}{F}_{h}+(N-3){G}_{h}{F}_{\iota} = 0 \nonumber\\
\iota&=&{G}_{g}{F}_{\iota}+4{G}_{h}{F}_{h}+2(N-4){G}_{h}{F}_{\iota} = 0.
\nonumber
\end{eqnarray}
\noindent which gives for $\eta_0,\, \Delta_0, \,\eta_1, \, \Delta_1$:
\begin{equation}
\label{eq:G_HSapp}
\begin{array}{lll}
\eta_0 & = & 4 \nonumber \\
\Delta_0 & = & 16 \nonumber \\ 
\eta_1 & = & 4 \nonumber \\
\Delta_1 & = & 16
\end{array}
\end{equation}
\noindent yielding a value of $2$ for each frequency 
(See Eqs.~\ref{eq:omegasapp}.) in units of the trap 
frequency as expected.

\renewcommand{\theequation}{G\arabic{equation}}
\setcounter{equation}{0}
\section{Limits for the analytic, angular frequencies for large values of $\bar{V}_0$.}\label{app:limits}

Now consider the strength of the 
interparticle
interaction, $\bar{V}_0$, increasing from the weak interactions of the BCS 
regime
to the strong interactions of unitarity. In the expanded view in the region 
of weak interactions in
Figs.~(\ref{fig:trialthirteenex})-(\ref{fig:trialfifteenex}), one can
see these angular frequencies, $\omega_{{0}^+}$, $\omega_{{1}^+}$, and 
$\omega_{2}$, separate as the interaction gets stronger
and in Figs.~(\ref{fig:trialthirteenU}) and (\ref{fig:trialfourteenU}) stabilize
at multiples of the trap frequency. 
What is happening in the analytic expressions as $\bar{V}_0$ is increasing
for fixed $N$ to yield these stable limits? 
 
In this Appendix, I will use the analytic expressions for these three
frequencies to derive these limits, working with the roots, 
$\lambda_{\alpha} = \omega^2_{\alpha}$
in order to avoid the square root signs in the formulas.
The three angular roots, $\lambda_{{0}^+}$, $\lambda_{{1}^+}$, and 
$\lambda_{2}$,
are given in terms of the $F$ and $G$ elements by:
\begin{eqnarray}\label{eq:lambdaFG}
\lambda_{{0}^+} &=&  \left[G_g+2(N-2)G_h\right]\times \nonumber \\
&& \left[F_g+2(N-2)F_h+\frac{(N-2)(N-3)}{2}F_{\iota}\right] \label{eq:lambda0pFG} \\ 
\lambda_{{1}^+} &=&  \left[G_g+(N-4)G_h\right]\times \nonumber \\
 && \left[F_g+(N-4)F_h-(N-3)F_{\iota}\right] \label{eq:lambda1pFG} \\
\lambda_2 &=&  \left[G_g-2G_h\right] \left[F_g-2F_h+F_{\iota}\right] \label{eq:lambda2FG}
\end{eqnarray} 
Using the definitions of $ G_g, G_h, F_g, F_h, \mbox{and} \, F_i$ 
found in Appendix~\ref{app:FG}, and letting $\bar{V}_0$ 
become large,
a little numerical work reveals that the dominant terms will involve powers
of $N\gamma_{\infty}$ which limits to a value of $-1$ as
 $\bar{V}_0 \rightarrow 1.0$.  Since $\gamma_{\infty} \rightarrow O(-1/N)$ 
for ensemble sizes relevant to experiment, extra factors of $\gamma_{\infty}$ in
a term will make it drop out.  These limits for $N\gamma_{\infty}$
and $\gamma_{\infty}$ are easily determined numerically 
and shown in Figs.~(\ref{fig:trialforty})-(\ref{fig:trialfortyone})
for two values of $N$. Similarly,
limits for $\tanh\Theta_{\infty}$ and $\mbox{sech}^2\Theta_{\infty}$ as 
$\bar{V}_0$ increases
to unitarity can also be obtained numerically.
\begin{figure}
\centering
\begin{subfigure}[h]{0.51\textwidth}
\includegraphics[scale=0.6]{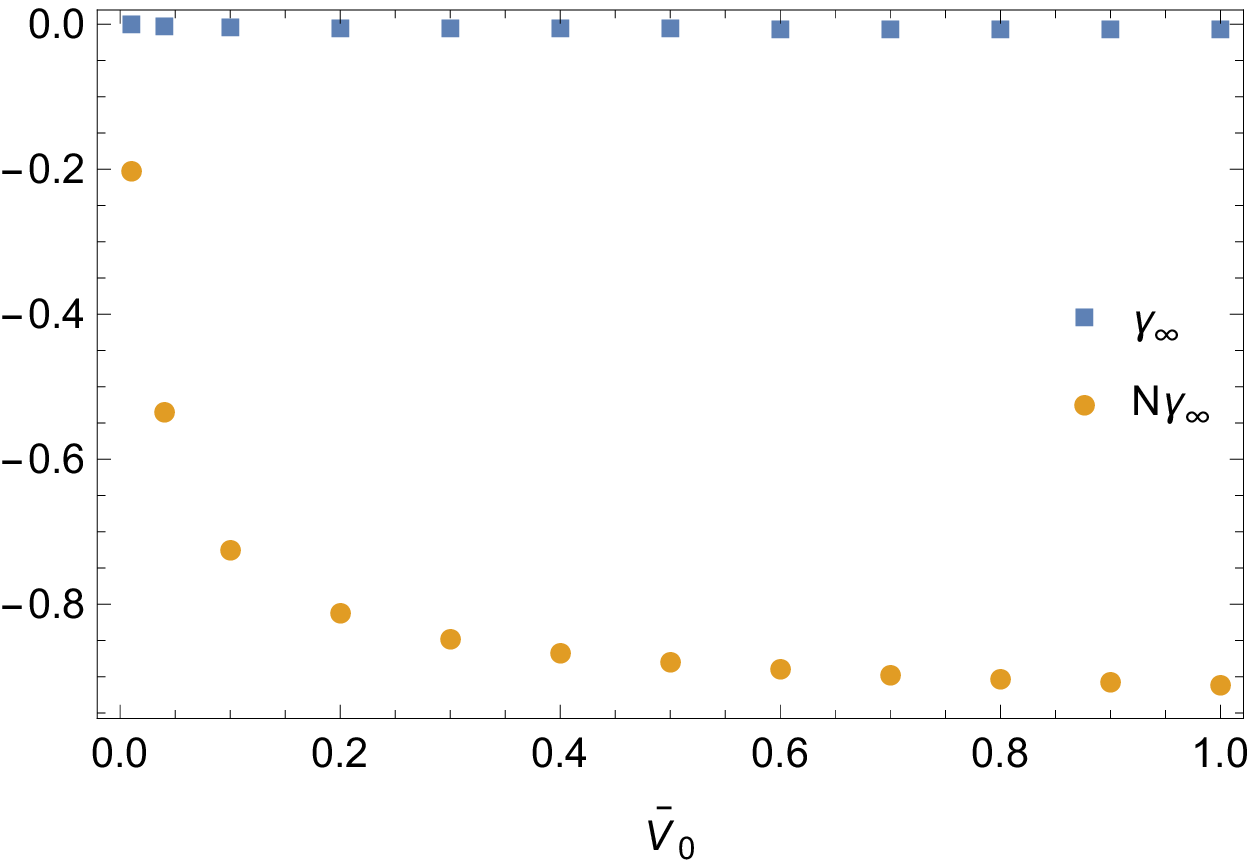}
\caption{a.$N=100$}
\label{fig:trialforty}
\end{subfigure}
\begin{subfigure}[h]{0.51\textwidth}
\includegraphics[scale=0.6]{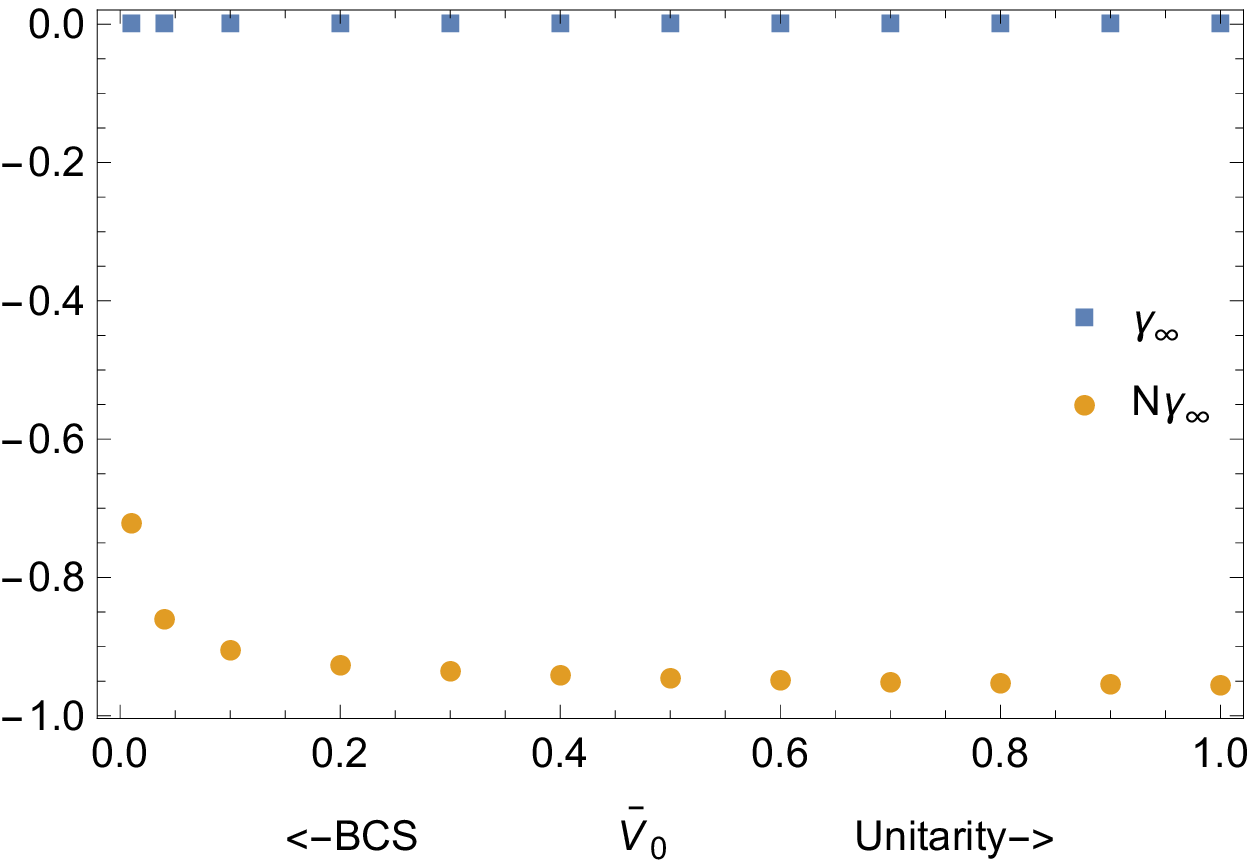}
\caption{b.$N=1000$} 
\label{fig:trialfortyone}
\end{subfigure}
\setcounter{figure}{5}
\caption{Limits  of $\gamma_{\infty}$ and $N\gamma_{\infty}$  as a function
of $\bar{V}_0$.}
\label{fig:forty2} 
\end{figure}
\begin{eqnarray}\label{eq:gammalimits}
\gamma_{\infty} & \rightarrow & O(-1/N) \nonumber \\
N\gamma_{\infty} & \rightarrow & -1. \nonumber \\
\mbox{sech}^2\Theta_{\infty} & \rightarrow & 0 \nonumber \\
\tanh\Theta_{\infty} & \rightarrow & 1, \nonumber
\end{eqnarray}
as $\bar{V}_0 \rightarrow 1.0$ 

Note that the powers of $N\gamma_{\infty}$ in the analytic expressions 
for the angular frequencies can begin to
dominate the expressions in two ways: 1) As $\bar{V}_0$
increases, the magnitude of $\gamma_{\infty}$ increases driving $N\gamma_{\infty}$
towards its limit of $-1$ as $N$ remains fixed. 2) Letting $N$ increase 
for a fixed value of $\bar{V}_0$, will also drive  
$N\gamma_{\infty}$
towards its limit of $-1$. This explains the earlier observation that:
``increasing the interaction between a fixed number of
 particles or increasing the number
of particles experiencing a fixed interaction has a similar effect in
separating the frequencies quickly.'' 
The microsopic dynamics
are however distinct between the two, 
as discussed in Section~\ref{subsec:microdiscussion}.

\noindent 
Using the above limits and the relation 
$\bar{r}_{\infty} = \frac{1}{\sqrt{2}\sqrt{1+(N-1)\gamma_{\infty}}}$,
in the definitions of $ G_g, \, G_h, \, F_g, \, F_h, \, \mbox{and} \, F_i$ 
in Appendix~\ref{app:FG} and keeping powers of $\gamma_{\infty}$ that
will contribute when factors of $(N-2)$, $(N-4)$ etc. are included from the 
expressions for the roots, $\lambda_{{0}^+}$, $\lambda_{{1}^+}$, and 
$\lambda_{2}$, the limits for these 
$G$ and $F$ elements 
for values of $N$
typical of experiments are:
%
\begin{eqnarray}
\label{eq:G_HSlimits}
{G}_{a}&=& \bm{1} \nonumber \\
{G}_{g}&=& 2\frac{1-{\gamma_{\infty}^2}}{{\bar{r}_{\infty}^2}} 
\rightarrow  \bm{\frac{2}{\bar{r}_{\infty}^2}} \nonumber \\
{G}_{h}&=& \frac{\gamma_{\infty}(1-\gamma_{\infty})}{{\bar{r}_{\infty}}^2} 
\approx  \bm{\frac{\gamma_{\infty}}{\bar{r}_{\infty}^2}}  
\end{eqnarray}
\begin{eqnarray}
\label{eq:F_HSlimits} 
{F}_{g}&=&\frac{1}{2\bar{r}_{\infty}^2(1-\gamma_{\infty})^3(1+(N-1)\gamma_{\infty})^{3}} \nonumber \\
&& \times\Bigl(1+3(N-2)\gamma_{\infty} + (13-11N+3N^2){\gamma_{\infty}}^2 \nonumber \\
 &\; \;& + (N-2)(4-3N+N^2){\gamma_{\infty}}^3\Bigr) + \frac{\bar{V}_{o} \bar{c}_{o}}{2}\mbox{sech}^2\Theta_{\infty} \nonumber \\
&& \times \Biggl[\frac{\bar{c}_{o}\bar{r}^2_{\infty}}{1-\gamma_{\infty}}
\tanh\Theta_{\infty} + \frac{\bar{r}_{\infty}}{2(1-\gamma_{\infty})^{3/2}}
\Biggr] \nonumber \\
& \rightarrow & \frac{1}{2\bar{r}_{\infty}^2(1-\gamma_{\infty})^3(1+(N-1)\gamma_{\infty})^{3}} \nonumber \\
&& \times\Bigl(1+3N\gamma_{\infty} + 3N^2{\gamma_{\infty}}^2 + N^3{\gamma_{\infty}}^3\Bigr) \nonumber \\
&=& \frac{1}{2\bar{r}_{\infty}^2(1-\gamma_{\infty})^3(1+(N-1)\gamma_{\infty})^{3}} \times (1+N\gamma_{\infty})^3 \nonumber \\
& \approx & \mathbf{1/(2\bar{r}_{\infty}^2)} \\
{F}_{h}&=&\frac{-\gamma_{\infty}}
{4\bar{r}_{\infty}^2(1-\gamma_{\infty})^3(1+(N-1)\gamma_{\infty})^{3}} \nonumber \\
&& \times \left[3+(5N-14)\gamma_{\infty} + (11-9N+2N^2){\gamma_{\infty}}^2
\right] \nonumber  \\
& \rightarrow &\frac{-\gamma_{\infty}}
{4\bar{r}_{\infty}^2(1-\gamma_{\infty})^3(1+(N-1)\gamma_{\infty})^{3}} \nonumber \\
&&\times \left[3+5N\gamma_{\infty} + 2N^2{\gamma_{\infty}}^2\right] \nonumber\\
&=&\frac{-\gamma_{\infty}}
{4\bar{r}_{\infty}^2(1-\gamma_{\infty})^3(1+(N-1)\gamma_{\infty})^{3}} \nonumber \\
&&\times \left[(3+2N\gamma_{\infty})(1 + N{\gamma_{\infty}})\right] \nonumber \\
& \approx & \frac{-\gamma_{\infty}}
{4\bar{r}_{\infty}^2(1-\gamma_{\infty})^3(1+(N-1)\gamma_{\infty})^{2}} \nonumber \\
& \approx & \bm{-\gamma_{\infty}\bar{r}_{\infty}^2}  \\
F_{\iota}&=& \frac{\gamma_{\infty}^2(2+(N-2)\gamma_{\infty})}{\bar{r}_{\infty}^2(1-\gamma_{\infty})^3(1+(N-1)\gamma_{\infty})^3} \nonumber \\
& \rightarrow & \frac{\gamma_{\infty}^2 8\bar{r}_{\infty}^6}{\bar{r}_{\infty}^2
(1-\gamma_{\infty})^3} = \frac{8\gamma_{\infty}^2 \bar{r}_{\infty}^4}{(1-\gamma_{\infty})^3} \nonumber \\
& \approx & \bm{8\gamma_{\infty}^2 \bar{r}_{\infty}^4}
\end{eqnarray}

Looking at the expressions for  the three angular roots, 
$\lambda_{{0}^+}$, $\lambda_{{1}^+}$, and 
$\lambda_{2}$, the following combinations are needed:
$G_g+2(N-2)G_h,\,\,
G_g+(N-4)G_h ,\,\,G_g-2G_h,\,\,2(N-2)F_h,\,\,\frac{(N-2)(N-3)}{2}F_i,\,\, 
(N-4)F_h,\,\,\mbox{and} \,\,(N-3)F_i$.
Taking these limits gives:
\begin{eqnarray}
G_g+2(N-2)G_h &=& 2\frac{1-{\gamma_{\infty}}^2}{{\bar{r}_{\infty}}^2} + 2(N-2) \frac{\gamma_{\infty}(1-\gamma_{\infty})}{\bar{r}_{\infty}^2} \nonumber \\
&=&\frac{2}{\bar{r}_{\infty}^2} (1-\gamma_{\infty})(1+(N-1)\gamma_{\infty}) \nonumber \\
&=& \frac{(1-\gamma_{\infty})}{\bar{r}_{\infty}^4} \approx 
\bm{\frac{1}{\bar{r}_{\infty}^4}} \nonumber \\
G_g+(N-4)G_h &=&  2\frac{1-{\gamma_{\infty}}^2}{{\bar{r}_{\infty}}^2} + (N-4)\frac{\gamma_{\infty}(1-\gamma_{\infty})}{{\bar{r}_{\infty}}^2} \nonumber \\
&=&\frac{(1-\gamma_{\infty})}{{\bar{r}_{\infty}}^2} (2+(N-2)\gamma_{\infty}) \nonumber \\
& \approx & \mathbf{\frac{1}{{\bar{r}_{\infty}}^2}} \nonumber \\
G_g-2G_h &=&  2\frac{1-{\gamma_{\infty}}^2}{{\bar{r}_{\infty}}^2} -2\frac{\gamma_{\infty}(1-\gamma_{\infty})}{{\bar{r}_{\infty}}^2} \nonumber \\
&=& \frac{2}{\bar{r}_{\infty}^2}\left[(1-\gamma_{\infty}^2)-\gamma_{\infty}(1-\gamma_{\infty})\right] \nonumber \\
&=& \frac{2(1-\gamma_{\infty})}{\bar{r}_{\infty}^2} \approx 
\bm{\frac{2}{\bar{r}_{\infty}^2}} \nonumber \\
2(N-2)F_h & \approx & 2(N-2)(-\gamma_{\infty}\bar{r}_{\infty}^2) \approx 
\bm{2\bar{r}_{\infty}^2} \nonumber \\
\frac{(N-2)(N-3)}{2} F_i &=& \frac{N^2-5N+6}{2} F_i \approx
\mathbf{4{\bar{r}_{\infty}^4}} \nonumber \\
(N-4)F_h &=& (N-4) (-\gamma_{\infty}\bar{r}_{\infty}^2) = 
\mathbf{{\bar{r}_{\infty}^2}} \nonumber \\
(N-3)F_i & \approx & (N-3)8\gamma_{\infty}^2 \bar{r}_{\infty}^4 \nonumber \\
& \approx &\bm{-8\gamma_{\infty} \bar{r}_{\infty}^4}  
\end{eqnarray} 

Substituting these results into the expressions for the roots,
Eqs.~(\ref{eq:lambda0pFG})-(\ref{eq:lambda2FG}) gives:
\begin{eqnarray}
\lambda_{{0}^+} &=& \frac{1}{\bar{r}^4_{\infty}}  \times \left[\frac{1}{2\bar{r}^2_{\infty}} + 2\bar{r}^2_{\infty} + 4\bar{r}^4_{\infty} \right] \nonumber \\
& \approx & 4 \label{eq:lambda0limit} \\
%
\lambda_{{1}^+} &=& \frac{1}{\bar{r}^2_{\infty}} \times \left[\frac{1}{2\bar{r}^2_{\infty}} + \bar{r}^2_{\infty} + 8\gamma_{\infty}\bar{r}^4_{\infty} \right] \nonumber \\
&=& \frac{1}{2\bar{r}^4_{\infty}} + 1 + 8\gamma_{\infty}\bar{r}^2_{\infty} 
\nonumber \\
& \approx & 1 \label{eq:lambda1limit} \\
%
\lambda_2 &=& \frac{2}{\bar{r}^2_{\infty}} \times \left[ \frac{1}{2\bar{r}^2_{\infty}} + 2\gamma_{\infty}\bar{r}^2_{\infty} 
+ 8\gamma^2_{\infty}\bar{r}^4_{\infty} \right] \nonumber \\
&=& \frac{1}{\bar{r}^4_{\infty}} + 4\gamma_{\infty}
+16\gamma^2_{\infty}\bar{r}^2_{\infty} \approx \frac{1}{\bar{r}^4_{\infty}}
\nonumber \\
& \approx &  = O(10^{-4}) \,\,\,\,\mbox{for the unitary regime},\label{eq:lambda2limit} 
\end{eqnarray}

\noindent yielding values for $\lambda_{{0}^+}, \lambda_{{1}^+},$ and $\lambda_2$ 
of:
\begin{eqnarray}\label{eq:omegaintegers}
\omega_{{0}^+} &=& 2 \nonumber \\
\omega_{{1}^+} &=& 1 \nonumber \\
\omega_2 &=& O(10^{-2}) 
\end{eqnarray}
\noindent in units of the trap frequency, $\omega_{ho}$.
Thus, as expected from physical arguments (see Section~\ref{sec:micro}.), 
these expressions for the angular frequencies limit to multiples of the 
trap frequency as $\bar{V}_0$ and/or $N$ increase.

\twocolumngrid


\renewcommand{\theequation}{B\arabic{equation}}
\setcounter{equation}{0}


\begin{thebibliography}{99}
\bibitem{jin1} M. Greiner, C.A. Regal, and D.S. Jin, Nature \textbf{426}, 
537(2003).
\bibitem{zwierlein2} M.W. Zwierlein, C.A. Stan, C.H. Schunck, S.M.F. Raupach, 
S. Gupta, Z. Hadzibabic, and W. Ketterle, Phys. Rev. Lett. \textbf{91}, 250401
(2003).
\bibitem{jochim1} S. Jochim, M. Bartenstein, A. Altmeyer, G. Hendl, 
S. Riedl, C. Chin, J. Hecker Denschlag, R. Grimm, Science \textbf{302}, 
2101 (2003).
\bibitem{thomas1} J. Kinast, S.L. Hemmer, M.E. Gehm, A. Turlapov and
J.E. Thomas, Phys. Rev. Lett. \textbf{92}, 150402 (2004).
\bibitem{salomon1} T. Bourdel, L. Khaykovich, J. Cubizolles, J. Zhang, F. Chevy, M. Teichmann, L. Tarruell, S.J.J.M.F. Kokkelmans and C. Salomon, Phys. Rev.
Lett. \textbf{93}, 050401 (2004).
\bibitem{jin2} C.A. Regal, M. Greiner, and D.S. Jin, Phys. Rev. Lett. 
\textbf{92}, 040403(2004).
\bibitem{zwierlein1}  M.W. Zwierlein, C.A. Stan,C.H. Schunck,S.M.F. Raupach, 
A.J. Kerman, and W. Ketterle, Phys. Rev. Lett. \textbf{92}, 120403(2004).
\bibitem{grimm1} C. Chin, M. Bartenstein, A. Altmeyer, S. Riedl, S. Jochim, 
Hecker-Denschlag and R. Grimm, Science \textbf{305}, 1128 (2004).

\bibitem{hulet1} G.B. Partridge, K.E. Strecker, R.I. Kamar, M.W. Jack and 
R.G. Hulet, Phys. Rev. Lett. \textbf{95}, 020404 (2005).
\bibitem{bcs} J. Bardeen, L.N. Cooper, and J.R. Schrieffer, Phys. Rev.
 \textbf{108}, 1175 (1957).
\bibitem{leggett1} A.J. Leggett, In {\it Modern Trends in the Theory of 
Condensed Matter. Proceedings of the XVIth Karpacz Winter School of
Theoretical Physics, Karpacz, Poland}, pgs. 13-27, Springer-Verlag, Berlin, 1980.
\bibitem{leggett2} A.J. Leggett, Rev. Mod. Phys. \textbf{73}, 307(2001).
\bibitem{eagles} D.M. Eagles, Phys. Rev. \textbf{186}, 456 (1969).
\bibitem{nozieres} P. Nozieres and S. Schmitt-Rink, J. Low Temp. Phys
.\textbf{59}, 195 (1985). 
\bibitem{giorgini1} S. Giorgini, L.P. Pitaevskii, and S. Stringari,
Rev. Mod. Phys. \textbf{80}, 1215(2008).
\bibitem{randeria1} M. Randeria and E. Taylor, Ann. Rev. Condensed Matter
Phys. \textbf{5}, 209(2014).

\bibitem{NM4} D.L. Rousseau, R.T. Bauman, S.P.S. Porto, J. Ramam Spect.,
\textbf{10}, 253(1981).
\bibitem{NM8} M.J. Clement, App J. \textbf{249}, 746(1981).
\bibitem{NM1} N. Zagar, J. Boyd, A. Kasaara, J. Tribbia, E. Kallen, H. Tanaka,
and J.-i Yano, Bull. Am. Meteor. Soc. \textbf{97} (2016).
\bibitem{NM2} S.C. Webb, Geophys.J. Int. \textbf{174}, 542(2008).
\bibitem{NM3} B.V. Sanchez, J. Marine Geodesy \textbf{31}, 181(2008).
\bibitem{NM5} J. Lee, K.T. Crampton, N. Tallarida, and V.A.Apkarian,
Nature \textbf{568}, 78(2019).
\bibitem{NM6} L. Fortunato, EPJ Web of Conferences \textbf{178}, 02017 (2018).
\bibitem{NM7} E.C. Dykeman and O.F. Sankey, J. Phys.:Condens. Matter \textbf{22}, 423202(2010).
\bibitem{NM9} M. Coughlin and J. Harms, arXiv:1406.1147v1 [gr-qc] (2014).
\bibitem{NM10} K. D. Kokkolas, Class. Quantum Grav. \textbf{8}, 2217 (1991).
\bibitem{NM11} R.M. Stratt, Acc. Chem. Res. \textbf{28}, 201(1995).
\bibitem{NM12} C.R. McDonald, G. Orlando, J.W. Abraham, D. Hochstuhl, 
M. Bonitz, and T. Brabec, Phys. Rev. Lett. \textbf{111}, 256801 (2013); 
F. Dalfove, 
S. Giorgini, L.P. Pitaevskii, and S. Stringari, Rev. Mod. Phys. 
\textbf{71}, 463(1999);
D. Jaksch, C. Bruder, J.I. Cirac, C.W. Gardiner, and P. Zoller, 
Phys. Rev. Lett. \textbf{81}, 3108(1998); H. Dong, W. Zhang, L. Zhou and Y. Ma,
Sci. Rep. \textbf{5}, 15848; doi: 10.1038/srep15848(2015).
\bibitem{NM13} H.C. Nagerl, C. Roos, H. Rohde, D. Leibfried, J. Eschner, 
F. Schmidt-Kaler and R. Blatt, Fortschr. Phys. \textbf{48}, 623 (2000).
\bibitem{annphys} D.K. Watson, Ann. Phys. \textbf{419}, 168219 (2020).
\bibitem{FGpaper} B.A.\ McKinney, M.\ Dunn, D.K.\ Watson, and J.G.\ Loeser,
  Ann.\ Phys.\ \textbf{310}, 56 (2003).
\bibitem{paperI} M.\ Dunn, D.K.\ Watson, and J.G.\ Loeser,
Ann.\ Phys.\ (NY), \textbf{321}, 1939 (2006).

\bibitem{energy} B.A.\ McKinney, M.\ Dunn, D.K.\ Watson,
  Phys.\ Rev.\ A \textbf{69}, 053611 (2004).
\bibitem{laingdensity} W.B. Laing, M. Dunn, and D.K. Watson, 
 Phys. Rev. A \textbf {74}, 063605 (2006).
\bibitem{JMPpaper} W.B. Laing, M. Dunn, and D.K. Watson, 
 J. of Math. Phys. \textbf{50}, 062105 (2009).

\bibitem{test}W.B. Laing, D.W. Kelle, M. Dunn, and D.K. Watson, J Phys A 
\textbf {42}, 205307 (2009).
\bibitem{toth} M. Dunn, W.B. Laing, D. Toth, and D.K. Watson, 
 Phys. Rev A \textbf {80}, 062108 (2009).
\bibitem{prl}D.K. Watson, Phys. Rev. A \textbf{92}, 013628
(2015).


\bibitem{harmoniumpra} D.K. Watson, Phys. Rev. A \textbf{93}, 023622 (2016).
\bibitem{partition} D.K. Watson, Phys. Rev. A \textbf{96}, 033601(2017).
\bibitem{emergence} D.K. Watson, J. Phys. B. \textbf{52}, 205301 (2019).


\bibitem{rearrangeprl} D.K. Watson and M. Dunn,  Phys. Rev. Lett. 
\textbf {105}, 020402 (2010).
\bibitem{complexity} D.K. Watson and M. Dunn, J. Phys. B \textbf{45},
095002 (2012).

\bibitem{epaps} W.B. Laing, M. Dunn, and D.K. Watson, EPAPS Document Number
E-JMAPAQ-50-031904.
\bibitem{avery} J.\ Avery, D.Z.\ Goodson, D.R.\ Herschbach,
Theor.\ Chim.\ Acta \textbf{81}, 1 (1991).
\bibitem{dcw} E.B.\ Wilson, Jr., J.C.\ Decius, P.C.\ Cross,
\textit{Molecular vibrations: The theory of infrared and raman
vibrational spectra}. McGraw-Hill, New York, 1955.
\bibitem{hamermesh} M.\ Hamermesh, \textit{Group theory and its
application to physical problems}, Addison-Wesley, Reading, MA,
1962.
\bibitem{WDC} See for example Ref.~\cite{dcw}, Appendix XII, p.\ 347.
\bibitem{loeser} J.G.\ Loeser, J.\ Chem.\ Phys.\ \textbf{86}, 5635 (1987).
\bibitem{bartenstein} M. Bartenstein, A. Altmeyer, S. Riedl, S. Jochim,
C. Chin, J. H. Denschlag, and R. Grimm, Phys. Rev. Lett. \textbf{92}, 203201-1
(2004).
\bibitem{baranov} M.A. Baranov and D.S. Petrov, Phys. Rev. A \textbf{62}, 
041601(R) (2000).
\bibitem{herschbach} \textit{Dimensional Scaling in Chemical Physics}, edited by
D.R. Herschbach, J. Avery, and O. Goscinski (Kluwer, Dordrecht, 1992).
\bibitem{zhen} Z. Zhen and J. Loeser, in \textit{Dimensional Scaling in Chemical Physics}, edited by
D.R. Herschbach, J. Avery, and O. Goscinski, Chapter 3, p. 90 (Kluwer, Dordrecht, 1992).
\bibitem{kais1} S. Kais and D.R. Herschbach, J. Chem. Phys. \textbf{100},
4367 (1994).
\bibitem{kais2} S. Kais and R. Bleil, J. Chem. Phys. \textbf{102},
7472 (1995).
\end{thebibliography}
\end{document}